\documentclass[11pt,a4paper]{article}

\usepackage[utf8]{inputenc}
\usepackage[T1]{fontenc}
\usepackage[english]{babel} 
\usepackage{amsmath, amssymb, amsfonts} 
\usepackage{graphicx}
\usepackage{booktabs} 
\usepackage{caption}
\usepackage{subcaption} 
\usepackage{geometry}
\usepackage{hyperref} 
\usepackage{cite}

\usepackage{orcidlink} 
\usepackage{authblk}
\hypersetup{
    colorlinks=true,
    linkcolor=blue,
    filecolor=magenta,      
    urlcolor=cyan,
    citecolor=blue,
}

\title{Extended Phase Space Thermodynamics and Joule-Thomson Expansion of Regular AdS Black Holes in a String Cloud}

\author{
    Y.~Elaima\textsuperscript{1}\,\orcidlink{0009-0008-0280-1019}\thanks{E-mail: \texttt{y.elaima.ced@uca.ac.ma}}, 
    A.~Daassou\textsuperscript{1}\,\orcidlink{0000-0001-9439-5047}\thanks{E-mail: \texttt{a.daassou@uca.ac.ma}}, 
    H.~Lekbich\textsuperscript{1,2}\,\orcidlink{0009-0009-6079-0394}\thanks{E-mail: \texttt{h.lekbich@edu.umi.ac.ma}}, 
    F.~Oubbad\textsuperscript{1}\,\orcidlink{0009-0008-7971-2153}\thanks{E-mail: \texttt{f.oubbad.ced@uca.ac.ma}}\\
    \small $^{1}$ \textit{Laboratory of Physics, Energy, Environment, and Applications (LP2EA),} \\
    \small \textit{Polydisciplinary Faculty, Cadi Ayyad University, Sidi Bouzid, P.O.Box 4162, Safi, Morocco} \\
    \small $^{2}$ \textit{Modern Physics, Radiation and Applications team, SIMED Laboratory,} \\
    \small \textit{FST Errachidia, Moulay Ismail University of Meknes, Meknes, Morocco}
}

\date{\today}

\begin{document}

\maketitle

\begin{abstract}
We investigate the extended phase space thermodynamics of a static, spherically symmetric regular black hole immersed in a string cloud background. By identifying the negative cosmological constant as a thermodynamic pressure, our analysis indicates that the quantum-inspired regularizing core and the string cloud density significantly affect the thermal stability of the black hole. We specifically illustrate a stringent scale invariance linked to the regularizing core, while the dimensionless string cloud parameter substantially alters the universal Van der Waals critical compressibility ratio. The significant geometric inconsistencies result in computed critical exponents that align with those of mean-field theory. We specify the Joule-Thomson expansion to define the isenthalpic heating and cooling regimes, so providing a comprehensive thermodynamic characterization of this non-singular spacetime.
\end{abstract}
\vspace{1em}
\section{Introduction}
\label{sec:intro}

The thermodynamics of black holes has been a very fertile area for theoretical physics since the pioneering work by Bekenstein and Hawking \cite{Bekenstein1973, Hawking1975} that black holes are not only macroscopic geometric solutions of general relativity but also complex thermodynamic systems with temperature and entropy. The study of black holes in Anti-de Sitter (AdS) space has further enriched this deep connection between gravity, quantum mechanics and thermodynamics, where the presence of a negative cosmological constant provides a natural confining box allowing for stable thermodynamic equilibrium \cite{Chamblin1999}. 

The construction of the extended phase space thermodynamics \cite{Kastor2009, Dolan2011, Cvetic2011} was a major development in the thermodynamic interpretation of AdS black holes. In this novel framework the cosmological constant $\Lambda$ is not a fixed parameter anymore, but a dynamical thermodynamic pressure, $P = -\Lambda/8\pi$. It was shown that there is an exact correspondence between the phase transitions of charged AdS black holes and the liquid-gas phase transitions of the conventional Van der Waals (VdW) fluids \cite{Kubiznak2012}. This analogy involves first order phase transition, $P-V$ criticality and characteristic critical exponents, which are further generalized to rotating black holes, higher derivative gravity, nonlinear electrodynamics and even to cosmological de-Sitter backgrounds \cite{Gunasekaran2012, Altamirano2014, Hendi2017, Banerjee2012, Majhi2017, Li:2016zca}.

Despite these significant accomplishments, standard general relativity is nevertheless afflicted by the unavoidable occurrence of spacetime singularities places where curvature invariants diverge and the predictive capacity of physical laws breaks down \cite{Poisson1990}. To address this open issue without having to resort to a complete theory of quantum gravity, the idea of “regular black holes” was proposed. These models, pioneered by Bardeen \cite{Bardeen1968} and extended by Hayward \cite{Hayward2006}, remove the central singularity by means of a de Sitter core, usually associated to vacuum polarization or short-distance quantum gravitational effects \cite{Ansoldi2008, Balart2014, Fan2016}. The research on regular black holes has recently received a lot of interest \cite{Lan:2023cvz}, in particular their exact constructions \cite{Singh:2022xgi} and their complicated thermodynamic phase structures in AdS space \cite{Ma:2014qma, Singh:2020xju, Rehan:2024dsg, Singh:2024jgo}.

At the same time the macroscopic environment in the neighborhood of a black hole plays an important role in changing its shape and evolutionary dynamics. The so-called ``string cloud'' model first presented by Letelier \cite{Letelier1979} is a physically appealing astrophysical and cosmic background. The present model is inspired by string theory, and treats the universe as being populated by a fluid of one-dimensional extended objects – strings – rather than zero-dimensional point particles. The presence of a string cloud background introduces a dimensionless density parameter $\epsilon$ in the spacetime metric that gives rise to a strong modification of the horizon structure, the shadow and the global thermodynamic properties of the central black hole \cite{Ghaderi2016, Toledo2020, Morais2024, Ma2016, Santos2022}. A lot of theoretical works have been done recently on the interaction of ordinary black hole geometries with string cloud or fluid backgrounds, which have shown new phase transition dynamics \cite{Muniz:2025ugk, Mishra:2026iwq, Singh:2025svv,Daassou:2023tmp}.

Besides phase change, the Joule-Thomson (JT) expansion has become an important instrument to explore the thermodynamic property of AdS black hole, which gives a significant connection to the classical thermodynamics. The JT expansion describes in classical thermodynamics the change of temperature of a real gas during an isenthalpic expansion process. The enthalpy in the expanded phase space is associated with the black hole mass, and the JT expansion can be readily adapted to the AdS black holes. The JT expansion was first studied by \"{O}kc\"{u} and Ayd{\i}ner for charged AdS black holes \cite{Okcu2017} and has been extensively investigated for different black hole configurations, identifying the isenthalpic heating and cooling regions and the corresponding inversion curves \cite{Mo2018, Yerra2018, Spallucci2013,Lekbich:2023aop}. Recently, these techniques have been ingeniously extended to regular geometries, such as Bardeen-AdS spacetimes \cite{Li:2019jcd}. 

This study seeks to do a thorough examination of the extended phase space thermodynamics and the Joule-Thomson expansion of a static, spherically symmetric regular black hole situated within a string cloud environment, driven by these interrelated breakthroughs. In a recent complementary study \cite{Elaima2026}, we successfully constructed the rotating counterpart of this geometry a regular Kerr-like black hole in a string cloud and investigated its shadow and Hawking evaporation. We now concentrate on the precise thermodynamic phase structure, examining how the cohabitation of the quantum-inspired regularizing core ($r_0$) and the macroscopic string cloud ($\epsilon$) affects the $P-V$ criticality, the universal compressibility ratio, and thermal stability. Additionally, we offer a comprehensive account of the Joule-Thomson expansion for this non-singular spacetime.

The paper is structured as follows: In Section \ref{sec:exact_solution}, we introduce the metric of the regular black hole with a string cloud and analyze its horizon structure. Section \ref{sec:thermodynamics} is devoted to deriving the fundamental thermodynamic quantities and formulating the extended first law. In Section \ref{sec:equation_of_state}, we derive the equation of state and thoroughly examine the $P-V$ criticality and scale invariance. Section \ref{sec:stability} assesses the global and local stability of the system. In Section \ref{sec:jt_expansion}, we detail the Joule-Thomson expansion, deriving the inversion temperatures and curves. Finally, Section \ref{sec:conclusion} provides our concluding remarks.

Throughout this work, we adopt natural units where $G = \hbar = c = k_B=l_p = 1$.

\section{Exact Solution of the Regular AdS Black Hole with a Cloud of Strings}
\label{sec:exact_solution}

In this section, we establish the gravitational field equations and the matter sources that generate our regular black hole in Anti-de Sitter (AdS) space surrounded by a cloud of strings (CS) . Following the standard approach for exact black hole solutions \cite{Letelier1979, Fan2016}. In this framework, the short-distance regularizing core is modeled by an effective anisotropic energy-momentum tensor representing quantum-gravitational vacuum polarization or non-commutative geometry corrections \cite{Santos2022}, minimally coupled to gravity and a classical cloud of strings background.

The action of the system is governed by the Einstein-Hilbert term with a negative cosmological constant coupled to the effective matter fields:
\begin{equation}
    \mathcal{I} = \int d^4x \sqrt{-g} \left( \frac{R - 2\Lambda}{16\pi} \right) + \mathcal{I}_{\rm CS} + \mathcal{I}_{\rm eff},
\end{equation}
 where $g$ is the determinant of the metric tensor $g_{\mu\nu}$, $R$ is the Ricci scalar, and $\Lambda = -8\pi P$ is the negative cosmological constant identified with the thermodynamic pressure $P$ \cite{Kastor2009, Dolan2011, Kubiznak2012}. Furthermore, $\mathcal{I}_{\rm CS}$ denotes the action associated with the classical cloud of strings background, and $\mathcal{I}_{\rm eff}$ represents the effective action of the phenomenological matter fields responsible for generating the regular, non-singular core.

Varying the action with respect to the metric tensor yields the Einstein field equations:
\begin{equation}
    R_{\mu\nu} - \frac{1}{2}g_{\mu\nu}R + \Lambda g_{\mu\nu} = 8\pi T_{\mu\nu},
\end{equation}
where the total energy-momentum tensor is the sum of the cloud of strings (CS) and the effective quantum-inspired anisotropic fluid contributions:
\begin{equation}
    T_{\mu\nu} = T_{\mu\nu}^{\rm (CS)} + T_{\mu\nu}^{\rm (eff)}.
\end{equation}

The static, spherically symmetric cloud of strings yields a highly anisotropic energy-momentum tensor whose non-vanishing components are given by \cite{Letelier1979}:
\begin{equation}
    T^{t}_{\; t \, \rm (CS)} = T^{r}_{\; r \, \rm (CS)} = -\frac{\epsilon}{8\pi r^2}, \quad T^{\theta}_{\; \theta \, \rm (CS)} = T^{\phi}_{\; \phi \, \rm (CS)} = 0,
\end{equation}
where $\epsilon$ is the dimensionless cloud of strings parameter $(0 \le \epsilon < 1)$.

To find the black hole geometry, we consider a general static and spherically symmetric line element:
\begin{equation}
    ds^2 = -f(r)dt^2 + \frac{1}{f(r)}dr^2 + r^2(d\theta^2 + \sin^2\theta d\phi^2),
\end{equation}
where the metric function $f(r)$ is parameterized by an interpolating mass function $\mathcal{M}(r)$:
\begin{equation}
    f(r) = 1 - \frac{2\mathcal{M}(r)}{r} + \frac{8\pi P}{3}r^2.
\end{equation}

For this class of spherically symmetric spacetimes with $g_{tt} = -1/g_{rr}$, the effective matter source is described by an anisotropic fluid tensor $T^{\mu}_{\; \nu \, \rm (eff)} = \text{diag}(-\rho_{\rm eff}, p_r, p_\theta, p_\theta)$ that satisfies the radial equation of state $p_r = -\rho_{\rm eff}$.

To ensure a regular core at the origin ($r \to 0$) while recovering the Letelier-AdS geometry at spatial infinity, we prescribe the effective energy density profile as:
\begin{equation}
\label{eq:rho_eff}
    \rho_{\rm eff}(r) = \frac{e^{-r^3/r_0^3}}{8\pi} \left[ \frac{6M}{r_0^3} + \frac{3\epsilon r}{r_0^3} - \frac{\epsilon}{r^2} \right],
\end{equation}
where $M$ is the mass parameter and $r_0$ is the regularizing length scale. 

From the Einstein equations, the mass function $M(r)$ is governed by the radial differential equation:
\begin{equation}
    \mathcal{M}'(r) = 4\pi r^2 \rho_{\rm total}(r) = 4\pi r^2 \left( \rho_{\rm eff}(r) + \frac{\epsilon}{8\pi r^2} \right) = 4\pi r^2 \rho_{\rm eff}(r) + \frac{\epsilon}{2}.
\end{equation}
Integrating this equation from $0$ to $r$ with the central regularity condition $\mathcal{M}(0) = 0$ yields:
\begin{equation}
  \mathcal{M}(r) = \int_0^r \left( \frac{r^2 e^{-r^3/r_0^3}}{2} \left[ \frac{6M}{r_0^3} + \frac{3\epsilon r}{r_0^3} - \frac{\epsilon}{r^2} \right] + \frac{\epsilon}{2} \right) dr.
\end{equation}
Performing the integration, we analytically obtain:
\begin{equation}
    \mathcal{M}(r) = \left( M + \frac{\epsilon r}{2} \right) \left(1 - e^{-r^3/r_0^3}\right).
\end{equation}
Substituting $\mathcal{M}(r)$ back into the metric function yields the exact regular AdS black hole metric:
\begin{equation}
\label{eq:metric_function}
    f(r) = 1 - \left( \frac{2M}{r} + \epsilon \right)\left( 1 - e^{-\frac{r^3}{r_0^3}} \right) + \frac{8\pi P}{3}r^2.
\end{equation}
Here, $M$ represents the parameter associated with the black hole mass. In the extended thermodynamic phase space, $M$ is strictly identified with the enthalpy $H$ of the system \cite{Kastor2009}. The parameter $\epsilon$ denotes the dimensionless density of the string cloud ($ 0 \le \epsilon < 1 $). The fundamental feature of this geometry is governed by the length-scale parameter $r_0$, which acts as a regularization factor avoiding the central singularity ($r_0 > 0$).

To rigorously assess the regularity of the core, we perform a Taylor expansion of the metric function near the origin ($r \to 0$):
\begin{equation}
    f(r) \approx 1 - \left( \frac{2M}{r_0^3} - \frac{8\pi P}{3} \right) r^2 - \frac{\epsilon}{r_0^3} r^3 + \mathcal{O}(r^5).
\end{equation}
Due to the presence of the cloud of strings coupled to the exponential regularizing factor, a non-smooth term proportional to $r^3$ (an odd power of the radial coordinate) appears. This indicates that the metric tensor is of class $\mathcal{C}^2$ rather than $\mathcal{C}^\infty$ at the origin. Nevertheless, we compute the limiting behavior of the main curvature invariants to prove the absence of any physical curvature singularity. The Ricci scalar $R$, the Ricci tensor squared $R_{\mu\nu}R^{\mu\nu}$, and the Kretschmann scalar $K = R^{\alpha\beta\gamma\delta}R_{\alpha\beta\gamma\delta}$ near $r \to 0$ converge to strictly finite values:
\begin{align}
    \lim_{r\to 0} R &= \frac{24M}{r_0^3} - 32\pi P, \\
    \lim_{r\to 0} R_{\mu\nu} R^{\mu\nu} &= 36 \left( \frac{2M}{r_0^3} - \frac{8\pi P}{3} \right)^2, \\
    \lim_{r\to 0} K &= 24 \left( \frac{2M}{r_0^3} - \frac{8\pi P}{3} \right)^2.
\end{align}
Since all these invariants remain finite and smooth as $r \to 0$, the spacetime possesses a well-defined non-singular geometric core.

We now analyze the local energy conditions of the total energy-momentum tensor $T^{\mu}_{\;\nu} = \text{diag}(-\rho_{\rm total}, p_r, p_\theta, p_\theta)$. The total energy density $\rho_{\rm total}$ is finite at the origin:
\begin{equation}
    \lim_{r\to 0} \rho_{\rm total}(r) = \frac{3M}{4\pi r_0^3}.
\end{equation}
However, the radial derivative of the total density near the origin is governed by:
\begin{equation}
    \frac{d\rho_{\rm total}}{dr} \approx \frac{\epsilon}{2\pi r_0^3} + \mathcal{O}(r^2).
\end{equation}
For any non-vanishing cloud of strings parameter ($\epsilon > 0$), we have $\frac{d\rho_{\rm total}}{dr} > 0$ near $r = 0$, meaning the energy density initially increases when moving away from the center before decaying at larger distances. 

From the conservation equation $\nabla_\mu T^{\mu\nu} = 0$, the tangential pressure is given by $p_\theta = -\rho_{\rm total} - \frac{r}{2}\rho_{\rm total}'$. The Null Energy Condition (NEC) in the tangential direction requires:
\begin{equation}
    \rho_{\rm total} + p_\theta = -\frac{r}{2}\frac{d\rho_{\rm total}}{dr} \ge 0.
\end{equation}
Near the origin, this yields:
\begin{equation}
    \rho_{\rm total} + p_\theta \approx -\frac{\epsilon r}{4\pi r_0^3} < 0 \quad (\text{for } r > 0).
\end{equation}
Consequently, the tangential NEC is locally violated in the vicinity of the de Sitter core whenever $\epsilon > 0$. This localized violation of the energy conditions is a standard and expected physical feature of regular spacetimes \cite{Zaslavskii2010}, as negative tangential pressures are required to provide the repulsive gravitational effect necessary to prevent the collapse into a singular point.

\subsection{Limiting Cases}
\label{subsec:limiting_cases}

To establish the physical consistency of our model, it is instructive to verify that the general metric function defined in Eq.~\eqref{eq:metric_function} correctly reduces to well-known spacetime geometries under specific limiting conditions of the parameters $\epsilon$ and $r_0$:

\begin{itemize}
    \item \textbf{Absence of the string cloud ($\epsilon \to 0$):} 
    When the spacetime is devoid of the string cloud background, the metric function reduces to a regular Schwarzschild-AdS black hole where the parameter $r_0$ still ensures a non-singular core.
    
    \item \textbf{Singular limit ($r_0 \to 0$):} 
    By taking the limit $r_0 \to 0$, the exponential term $e^{-r^3/r_0^3}$ vanishes for any $r > 0$. The metric function simplifies to exactly recover the Letelier-AdS black hole solution \cite{Letelier1979}.
    
    \item \textbf{Standard Schwarzschild-AdS limit ($r_0 \to 0$ and $\epsilon \to 0$):} 
    In the absence of both the string cloud and the regularizing core, we recover the standard Schwarzschild-AdS metric, the foundational geometry for P-V criticality studies \cite{Kubiznak2012}.
\end{itemize}

\subsection{Horizon Structure}
\label{sec:horizons}

The horizon structure of the black hole is fundamentally determined by the roots of the metric function, obtained by solving the condition $f(r_+) = 0$. Depending on the specific configuration of the mass $M$, the string cloud density $\epsilon$, the regularization parameter $r_0$, and the thermodynamic pressure $P$, the spacetime may exhibit an inner Cauchy horizon ($r_-$), an outer event horizon ($r_+$), a single degenerate horizon (extremal black hole), or no horizon at all \cite{Ansoldi2008, Poisson1990}.
To investigate the effects of the regularization parameter and the string cloud on the horizon radii, we set the mass parameter to $M=1.0$ and the pressure to $P=0.01$. The numerical solutions for the inner and outer horizons across several configurations are compiled in Table \ref{tab:horizons}.

\begin{table}[h!]
    \centering
    \renewcommand{\arraystretch}{1.3}
    \begin{tabular}{lcccc}
        \toprule
        \textbf{Configuration} & $\epsilon$ & $r_0$ & $r_-$  & $r_+$  \\
        \midrule
        Standard Schwarzschild-AdS & $0$ & $0$ & - & $1.63430$ \\
        Schwarzschild-AdS & $0$ & $0.5$ & $0.25952$ & $1.63430$ \\
        Schwarzschild-AdS + String Cloud & $0.3$ & $0$ & - & $1.95830$ \\
       Regular Schwarzschild-AdS + String Cloud & $0.3$ & $0.5$ & $0.25415$ & $1.95830$ \\
         Enhanced String Cloud & $0.9$ & $0.5$ & $0.24488$ & $2.74130$ \\
         Large Regularizing Core & $0.5$ & $1.0$ & $0.73025$ & $2.20460$ \\
        \bottomrule
    \end{tabular}
    \caption{Numerical values of the inner ($r_-$) and outer ($r_+$) horizons for fixed parameters $M=1.0$ and $P=0.01$.}
    \label{tab:horizons}
\end{table}

\begin{figure}[h!]
    \centering
    \includegraphics[width=0.85\textwidth]{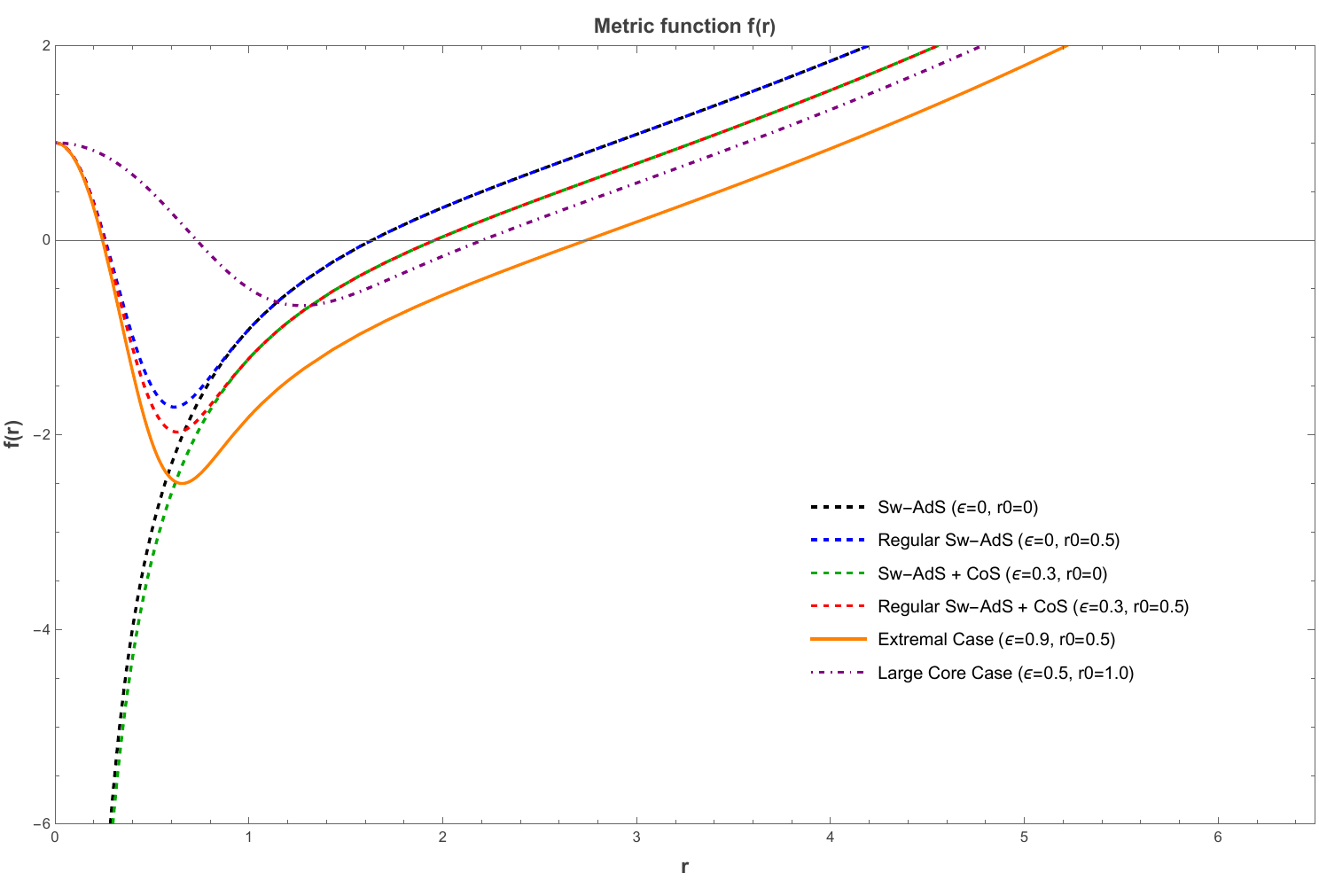}
    \caption{ Behavior of the metric function $f(r)$ with respect to the radial coordinate $r$ for $M=1.0$ and $P=0.01$. The intersections with the $f(r)=0$ axis denote the locations of the horizons $r_-$ and $r_+$.}
    \label{fig:metric_function}
\end{figure}
The metric function $f(r)$ is plotted in Fig.~\ref{fig:metric_function} for representative parameter choices at fixed  $M=1.0$ and $P=0.01$.
The graphical depiction of the metric function $f(r)$ for various scenarios is presented in Fig.~\ref{fig:metric_function}. The appearance of the Cauchy horizon is clear: when $r_0 = 0$, only an event horizon exists due to the singular core, whereas $r_0 > 0$ dynamically produces an inner horizon $r_-$. Augmenting $r_0$ displaces $r_-$ outward and $r_+$ inward, approaching an extremal condition. Moreover, the string density parameter $\epsilon$ displaces the outer event horizon $r_+$ outward, augmenting the black hole's effective size. This behavior is fully consistent with our previous study on the rotating counterpart of this solution, where the string cloud parameter was likewise shown to enlarge the event horizon and modify the spacetime geometry \cite{Elaima2026}.

\begin{figure}[h!]
    \centering
    \begin{subfigure}{0.48\textwidth}
        \includegraphics[width=\textwidth]{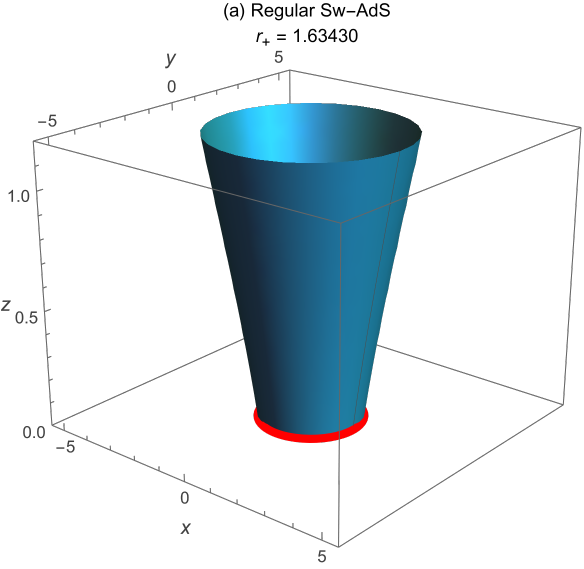}
        \caption{Regular Sw-AdS ($r_+=1.63430$)}
    \end{subfigure}
    \hfill
    \begin{subfigure}{0.48\textwidth}
        \includegraphics[width=\textwidth]{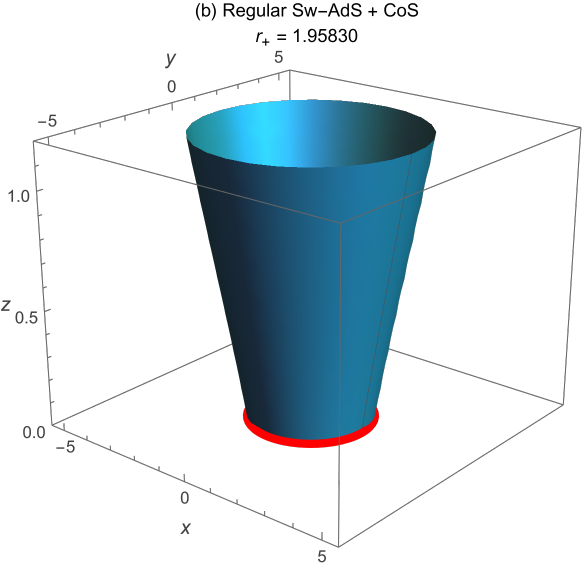}
        \caption{Regular Sw-AdS + CoS ($r_+=1.95830$)}
    \end{subfigure}
    \\ \vspace{0.5cm}
    \begin{subfigure}{0.48\textwidth}
        \includegraphics[width=\textwidth]{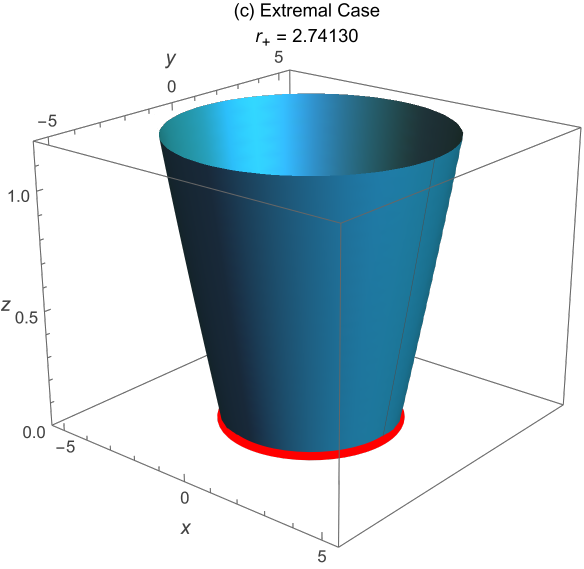}
        \caption{Enhanced String Cloud ($r_+=2.74130$)}
    \end{subfigure}
    \hfill
    \begin{subfigure}{0.48\textwidth}
        \includegraphics[width=\textwidth]{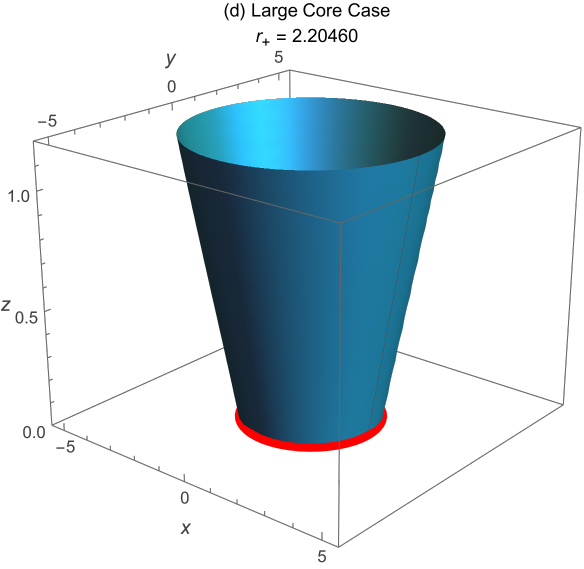}
        \caption{Large Core Case ($r_+=2.20460$)}
    \end{subfigure}
    \caption{3D spatial embeddings of the outer event horizon surface $r_+$ highlighting the geometrical deformation induced by the string cloud and the regularizing core parameters.}
    \label{fig:3d_horizons}
\end{figure}

The geometric deformation of the event horizon is further elucidated in Fig.~\ref{fig:3d_horizons}, confirming that both the quantum-inspired core and the string cloud profoundly alter the geometric cross-section of the black hole, which directly impacts its entropy and area-dependent thermodynamic properties \cite{Ansoldi2008, Morais2024}.

\section{Thermodynamic Properties}
\label{sec:thermodynamics}

To systematically study the thermal properties and ensure mathematical clarity, we introduce the auxiliary regularization function evaluated at the event horizon $r_+$:
\begin{equation}
    \Psi(r_+) = 1 - e^{-\frac{r_+^3}{r_0^3}}, \quad \text{with derivative} \quad \Psi'(r_+) = \frac{3r_+^2}{r_0^3} e^{-\frac{r_+^3}{r_0^3}}.
    \label{eq:psi_func}
\end{equation}
This function neatly isolates the effects of the non-singular core. In the singular limit ($r_0 \to 0$), $\Psi(r_+) \to 1$ and $\Psi'(r_+) \to 0$.

\subsection{Mass and Hawking Temperature}

The mass parameter $M$ of the black hole is obtained by solving the horizon condition $f(r_+) = 0$. Using our auxiliary function \eqref{eq:psi_func} yields :
\begin{equation}
    M = \frac{r_+}{2 \Psi(r_+)} \left[ 1 - \epsilon \Psi(r_+) + \frac{8\pi P}{3} r_+^2 \right].
    \label{eq:mass_psi}
\end{equation}

The Hawking temperature $T$ is derived from the surface gravity $\kappa$ via $T = \frac{f'(r_+)}{4\pi}$. By differentiating the metric function and replacing the mass parameter from Eq.~\eqref{eq:mass_psi}, the temperature is articulated as:
\begin{equation}
    T = \frac{1}{4\pi r_+} \left[ 1 - \epsilon \Psi(r_+) + 8\pi P r_+^2 - r_+ \frac{\Psi'(r_+)}{\Psi(r_+)} \left( 1 + \frac{8\pi P}{3} r_+^2 \right) \right].
    \label{eq:temperature_psi}
\end{equation}
To verify the physical consistency of our model, we examine the limit in which the regularizing core vanishes ($r_0 \to 0$) and the string cloud is absent ($\epsilon \to 0$). In this regime, the exponential term decays rapidly, yielding $\Psi(r_+) \to 1$ and $\Psi'(r_+) \to 0$. Consequently, the generalized Hawking temperature reduces perfectly to the well-known Schwarzschild-AdS expression, where the leading factor $1/(4\pi r_+)$ corresponds to the standard Schwarzschild background.

\begin{figure}[h!]
    \centering
    \includegraphics[width=0.75\textwidth]{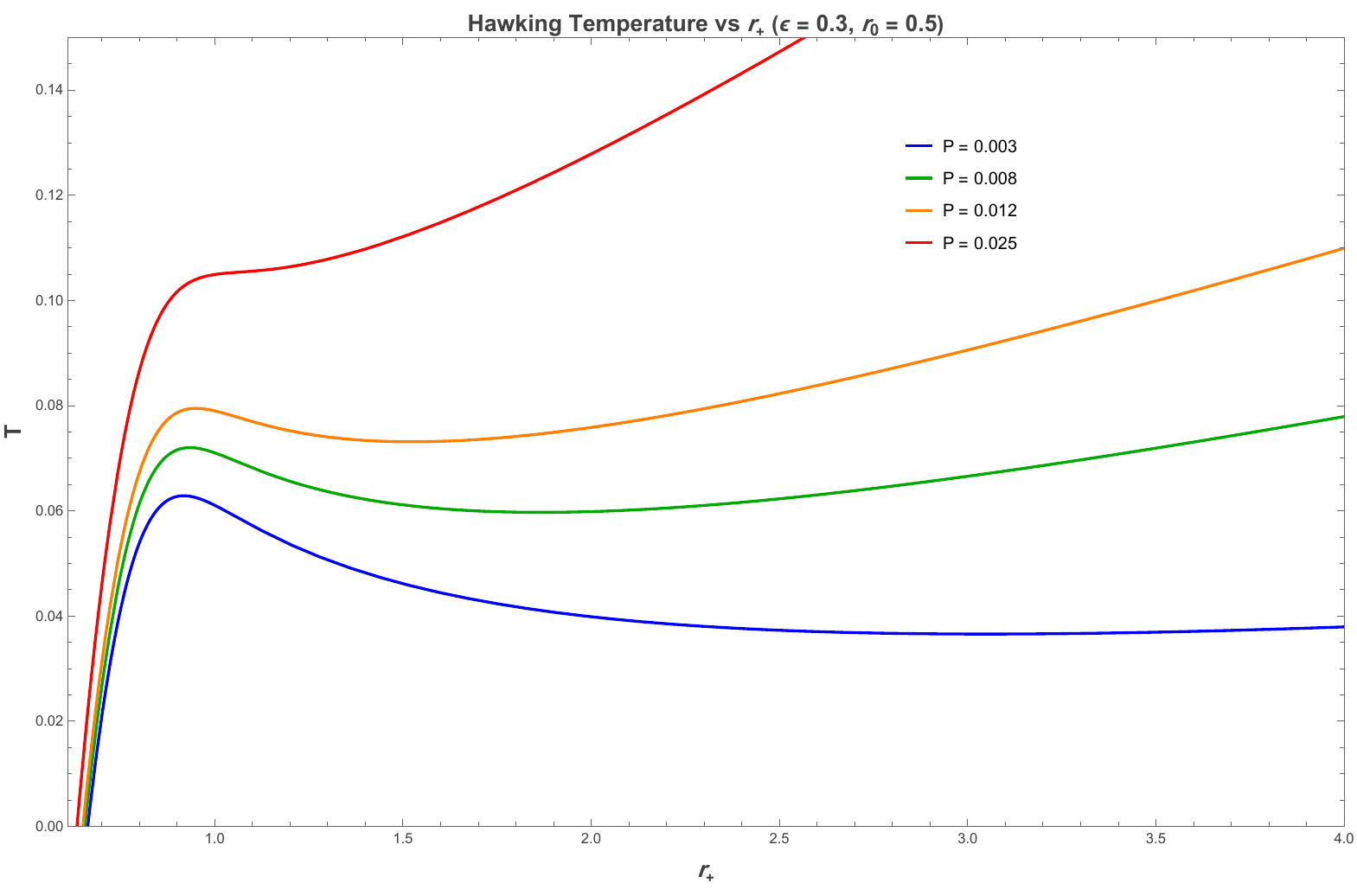}
    \caption{Hawking temperature $T$ as a function of $r_+$ for $\epsilon=0.3$ and $r_0=0.5$.}
    \label{fig:temperature}
\end{figure}

The Hawking temperature is depicted in Fig.~\ref{fig:temperature}. At low pressures, the temperature curve exhibits a local maximum and minimum, strongly suggesting a first-order phase transition \cite{Kubiznak2012, Wei2015}.

The thermodynamic behavior of the generalized Hawking temperature is depicted in Fig.~\ref{fig:temperature}. A highly prominent feature arising from the quantum-inspired regularizing length scale $r_0$ is observed in the small horizon regime ($r_+ \to 0$). Unlike the standard Schwarzschild-AdS black hole, where the temperature diverges to infinity as the horizon shrinks, our model exhibits a temperature that drops abruptly toward zero \cite{Elaima2026}. The intercept with the $r_+$-axis indicates the existence of an extremal black hole remnant with a non-zero horizon radius at absolute zero temperature, a direct consequence of the repulsive de Sitter-like core preventing total collapse. 

Furthermore, the temperature profiles strongly depend on the thermodynamic pressure $P$, displaying behavior perfectly analogous to a van der Waals fluid. At low, sub-critical pressures (e.g., $P = 0.003$ and $P = 0.008$), the isobaric curves exhibit a distinct local maximum followed by a local minimum, the intermediate branch represents a thermodynamically unstable phase, dictating that the system must undergo a first-order phase transition between the stable SBH and LBH configurations \cite{Kubiznak2012, Wei2015}.

As the pressure increases, the local extrema approach each other until they merge into a single inflection point at the critical pressure $P_c$ (approximated by the $P=0.012$ curve), characteristic of a second-order phase transition  \cite{Kubiznak2012}. For super-critical pressures (e.g., $P = 0.025$), the temperature increases monotonically with the horizon radius. In this regime, the distinction between small and large black holes vanishes, yielding a globally stable, continuous supercritical fluid-like phase. Finally, in the large horizon limit ($r_+ \to \infty$), the temperature increases linearly for all curves, driven by the dominant Anti-de Sitter cosmological pressure term $8\pi P r_+^2$.

\subsection{Entropy and Thermodynamic Volume}

For regular black holes, the entropy deviates from the Bekenstein-Hawking area law \cite{Fan2016, Hendi2017}. which states that the entropy is proportional to a quarter of the event horizon area $A_H$.
\begin{equation}
    S = \frac{A_H}{4} = \pi r_+^2 .
    \label{eq:entropy}
\end{equation}

In the extended phase space framework, the black hole mass $M$ represents the total enthalpy of the spacetime \cite{Kastor2009, Dolan2011}. Thus, the thermodynamic volume $V$, serving as the conjugate variable to the thermodynamic pressure $P$, is obtained by the conventional thermodynamic relation:
\begin{equation}
    V = \left( \frac{\partial M}{\partial P} \right)_{S, \epsilon, r_0}.
\end{equation}
Taking the derivative of the mass equation in Eq.~\eqref{eq:mass_psi} with respect to pressure results in the thermodynamic volume for our standard black hole model:
\begin{equation}
    V = \frac{4\pi r_+^3}{3 \Psi(r_+)} = \frac{4\pi r_+^3}{3 \left( 1 - e^{-\frac{r_+^3}{r_0^3}} \right)}.
    \label{eq:thermo_volume}
\end{equation}
In the limit of a singular black hole ($r_0 \to 0$, $\Psi \to 1$), we perfectly recover the standard geometric volume of a sphere $V = \frac{4\pi}{3} r_+^3$, which is strictly consistent with classical Schwarzschild-AdS black holes \cite{Cvetic2011, Altamirano2014}. However, due to the continuous mass distribution of the regularizing core, the thermodynamic volume is effectively inflated compared to the classical case. Most remarkably, in the limit where the black hole approaches complete evaporation and the horizon radius shrinks to zero ($r_+ \to 0$), the exponential term expands as $e^{-r_+^3/r_0^3} \approx r_+^3/r_0^3$. The thermodynamic volume then approaches a strictly finite minimal limit:
\begin{equation}
    \lim_{r_+ \to 0} V = \frac{4\pi}{3} r_0^3.
\end{equation}
This non-zero remnant volume serves as a direct macroscopic manifestation of the quantum-inspired core, providing a fundamental geometric buffer that strictly forbids the collapse of the spacetime into a point-like singularity \cite{Balart2014}.

\subsection{Extended First Law and Smarr Relation}

In the extended phase space framework, the black hole mass $M$ plays the role of thermodynamic enthalpy rather than internal energy. Consequently, the first law of thermodynamics must be generalized to incorporate the variations of all parameters governing the spacetime \cite{Kastor2009, Fan2016}, namely the pressure $P$, the string cloud density $\epsilon$, and the regularizing length scale $r_0$. The extended first law is formulated as:
\begin{equation}
    dM = T dS + V dP + \Phi d\epsilon + \mathcal{K} dr_0,
    \label{eq:first_law_extended}
\end{equation}
where $\Phi$ and $\mathcal{K}$ are the generalized thermodynamic potentials conjugate to the parameters $\epsilon$ and $r_0$, respectively. Because the entropy $S$ relies solely on $r_+$ and $r_0$ (see Eq.~\eqref{eq:entropy}), holding both $S$ and $r_0$ constant strictly implies holding the horizon radius $r_+$ constant. Thus, the thermodynamic potential $\Phi$ conjugate to the string cloud background is easily derived from Eq.~\eqref{eq:mass_psi}:
\begin{equation}
    \Phi = \left( \frac{\partial M}{\partial \epsilon} \right)_{S, P, r_0} = -\frac{r_+}{2}.
    \label{eq:phi_potential}
\end{equation}
Similarly, $\mathcal{K} = (\partial M / \partial r_0)_{S, P, \epsilon}$ represents the thermodynamic conjugate potential corresponding to the regularizing core, generally associated with the vacuum polarization effects that bypass the central singularity \cite{Balart2014, Hendi2017}.

To establish the corresponding Smarr relation, we utilize Euler's theorem for homogeneous functions \cite{Altamirano2014}. By designating the conventional geometric scaling dimensions: $[M]=L^1$, $[S]=L^2$, $[P]=L^{-2}$, and $[r_0]=L^1$. Importantly, the string cloud parameter $\epsilon$ denotes a dimensionless topological defect density, indicating that $[\epsilon] = L^0$ \cite{Letelier1979, Morais2024}. Utilizing Euler's scaling argument $M = \sum_i d_i X_i (\partial M / \partial X_i)$ inherently produces the generalized Smarr relation for our spacetime:
\begin{equation}
    M = 2TS - 2VP + \mathcal{K} r_0.
    \label{eq:smarr}
\end{equation}
The string cloud potential $\Phi$ is not explicitly incorporated in the Smarr formula. This preserves the standard thermodynamic scaling equilibrium while significantly modifying the equation of state, critical phase transitions, and free energy.

\section{P-V Criticality and the Equation of State}
\label{sec:equation_of_state}

To apply thermodynamics to a Van der Waals (VdW) fluid, it is imperative to derive the equation of state $P(V, T)$. By algebraically solving Eq.~\eqref{eq:temperature_psi} for the pressure $P$, we obtain the equation of state for the model:
\begin{equation}
\begin{split}
P(r_+, T) &= \frac{T}{2 r_+ \left( 1 - \frac{r_+ \Psi'(r_+)}{3\Psi(r_+)} \right)} - \frac{1}{8\pi r_+^2 \left( 1 - \frac{r_+ \Psi'(r_+)}{3\Psi(r_+)} \right)} \\
&\quad + \frac{\epsilon \Psi(r_+)}{8\pi r_+^2 \left( 1 - \frac{r_+ \Psi'(r_+)}{3\Psi(r_+)} \right)} + \frac{\Psi'(r_+)}{8\pi r_+ \Psi(r_+) \left( 1 - \frac{r_+ \Psi'(r_+)}{3\Psi(r_+)} \right)}.
\end{split}
\label{eq:eos_split}
\end{equation}

In standard extended phase space thermodynamics for AdS black holes, the specific volume is typically defined as \( v = 2r_+ \) \cite{Kubiznak2012}. However, the structure of Eq.~\eqref{eq:eos_split} reveals a universal modulating factor in the denominator. We therefore define the \textit{effective specific volume} $v_{\text{}}$ as:
\begin{equation}
    v_{\text{}} = 2 r_+ \left( 1 - \frac{r_+ \Psi'(r_+)}{3\Psi(r_+)} \right).
    \label{eq:veff}
\end{equation}
we can gracefully rewrite the expanded equation of state \eqref{eq:eos_split} into a compact, robust Van der Waals-like structure \cite{Gunasekaran2012, Hendi2017}:
\begin{equation}
    P(v_{\text{}}, T) = \frac{T}{v_{\text{}}} - \frac{1 - \epsilon \Psi(r_+) - r_+\frac{\Psi'(r_+)}{\Psi(r_+)}}{4\pi r_+ v_{\text{}}}.
    \label{eq:eos_vdw}
\end{equation}

\begin{figure}[h!]
    \centering
    \begin{subfigure}{0.48\textwidth}
        \includegraphics[width=\textwidth]{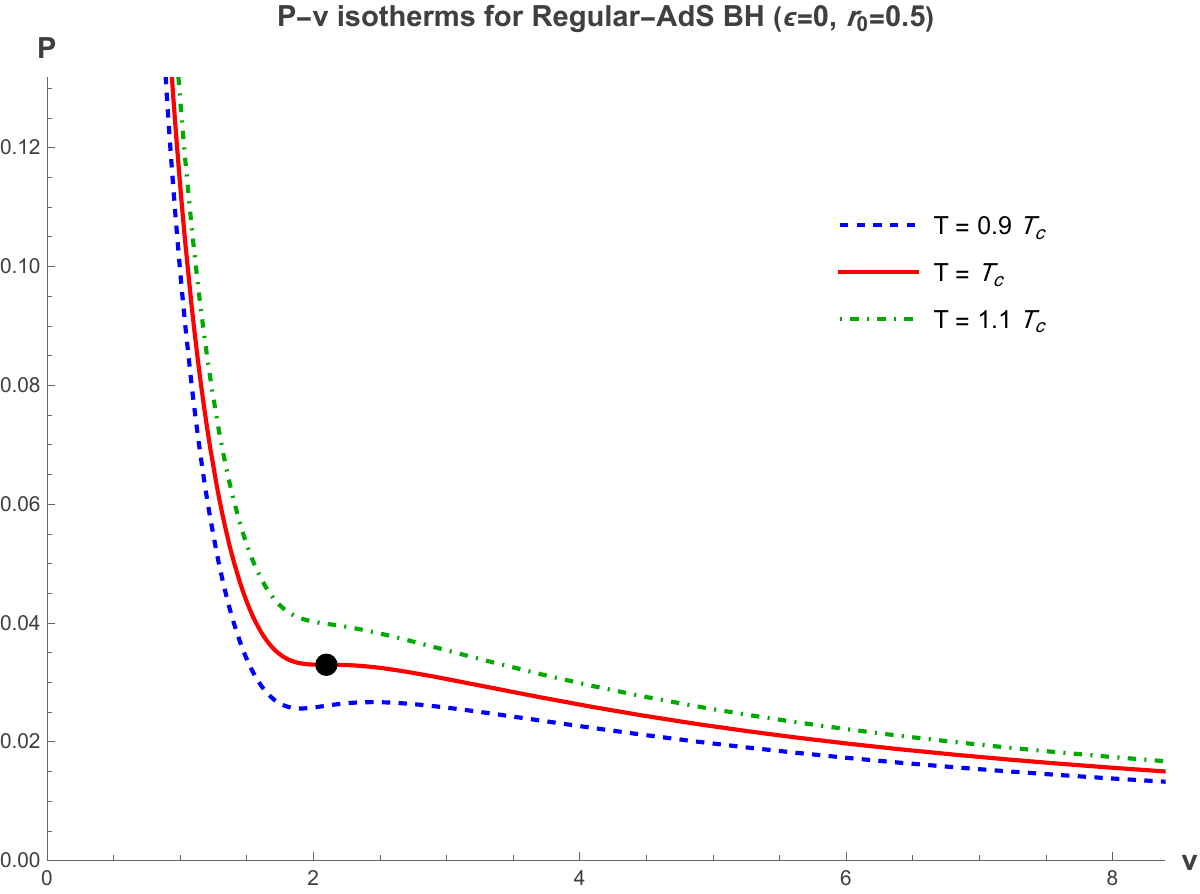}
        \caption{$\epsilon=0$}
    \end{subfigure}
    \hfill
    \begin{subfigure}{0.48\textwidth}
        \includegraphics[width=\textwidth]{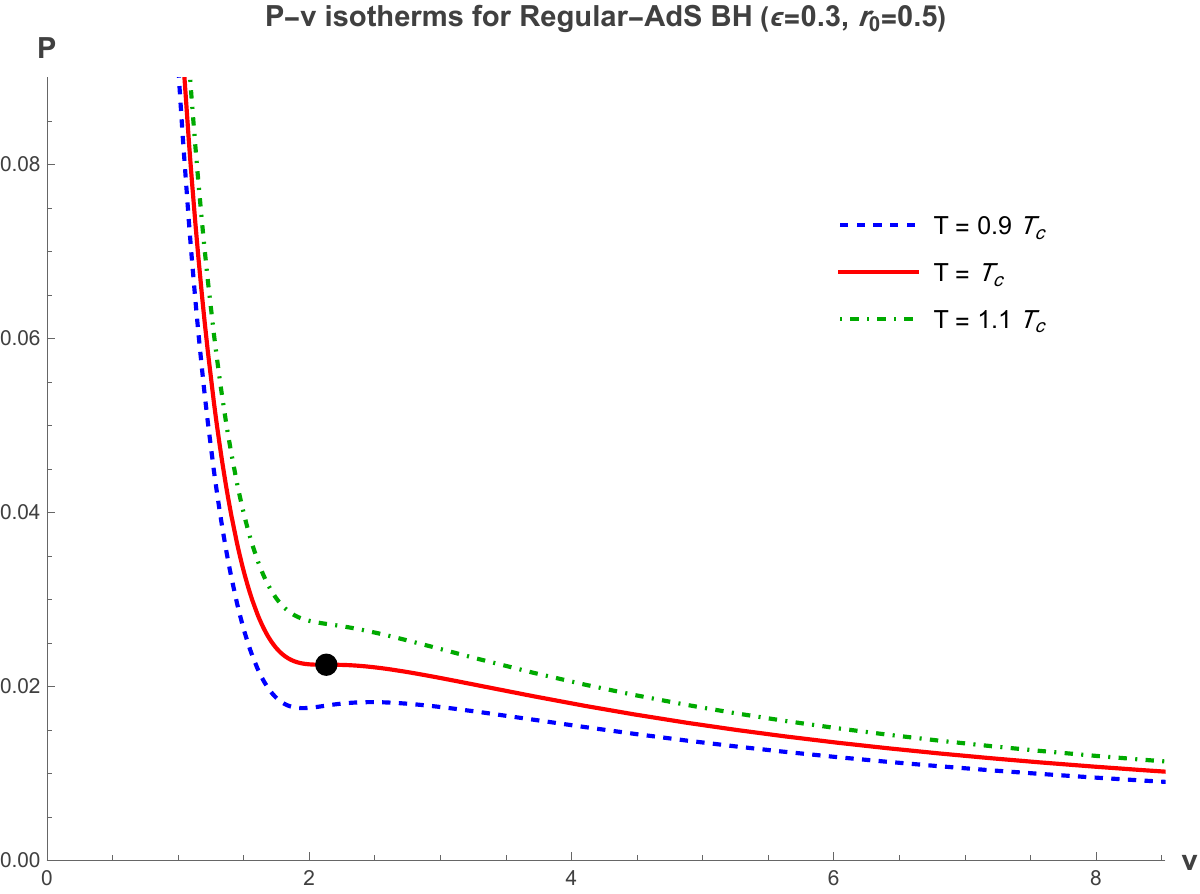}
        \caption{$\epsilon=0.3$}
    \end{subfigure}
    \\ \vspace{0.3cm}
    \begin{subfigure}{0.48\textwidth}
        \includegraphics[width=\textwidth]{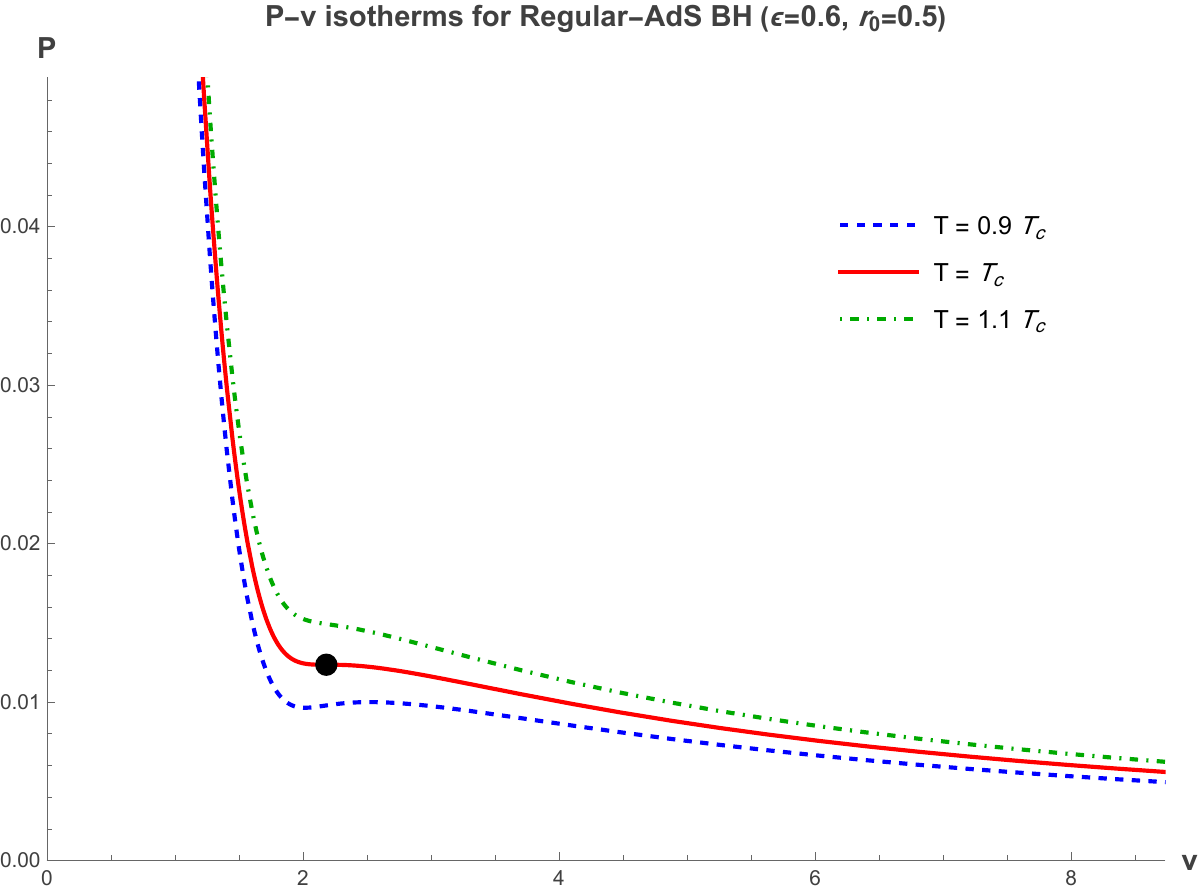}
        \caption{$\epsilon=0.6$}
    \end{subfigure}
    \hfill
    \begin{subfigure}{0.48\textwidth}
        \includegraphics[width=\textwidth]{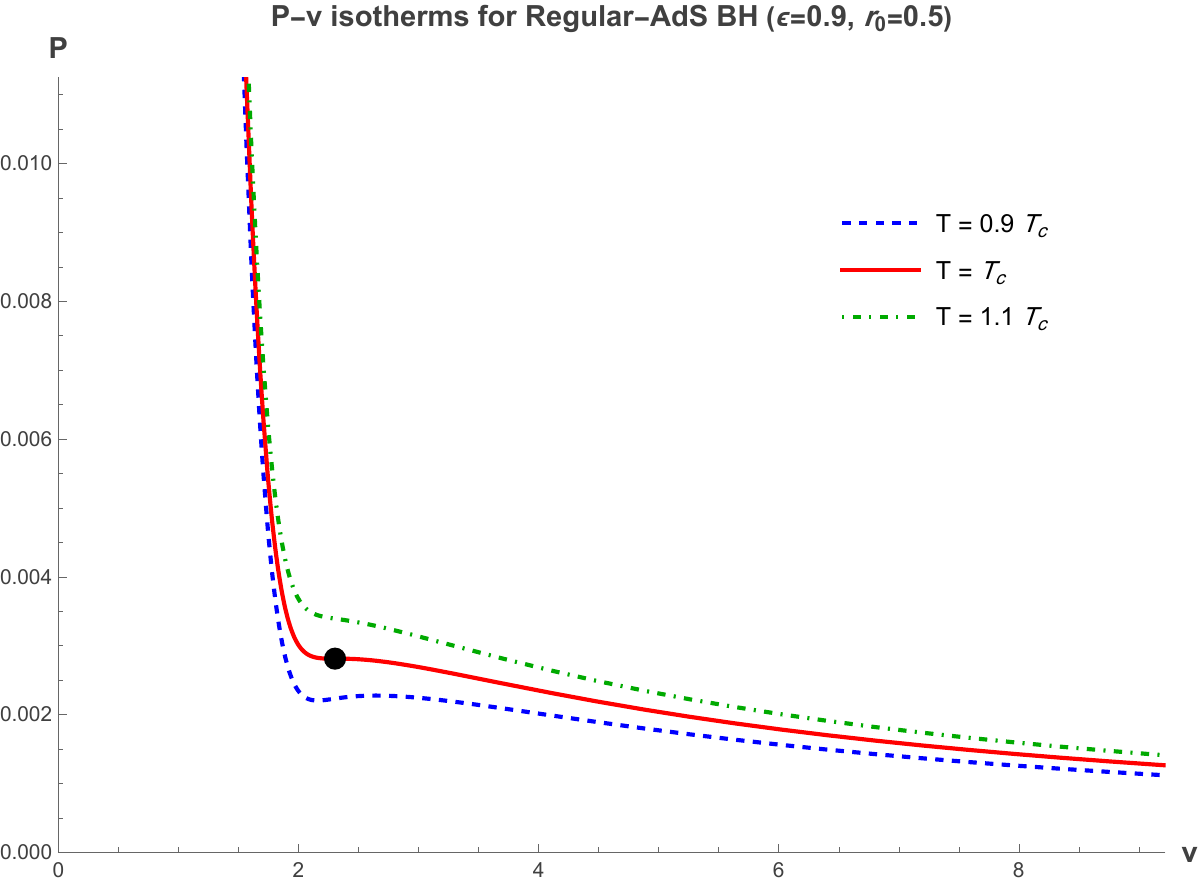}
        \caption{$\epsilon=0.9$}
    \end{subfigure}
    \caption{The $P-v$ isotherms for the regular black hole with fixed  $r_0=0.5$ and varying string cloud densities $\epsilon$.}
    \label{fig:pv_diagrams_eps}
\end{figure}

\begin{figure}[h!]
    \centering
    \begin{subfigure}{0.48\textwidth}
        \includegraphics[width=\textwidth]{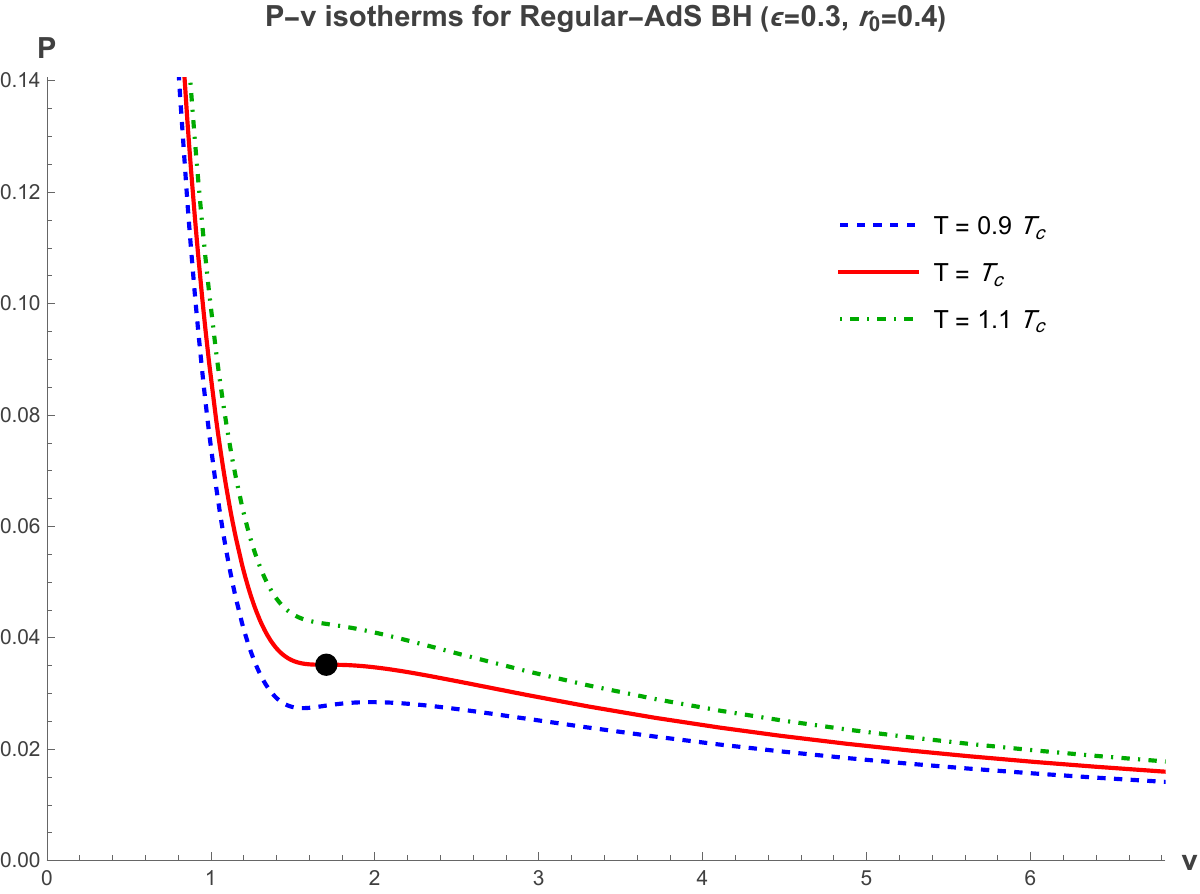}
        \caption{$r_0=0.4$}
    \end{subfigure}
    \hfill
    \begin{subfigure}{0.48\textwidth}
        \includegraphics[width=\textwidth]{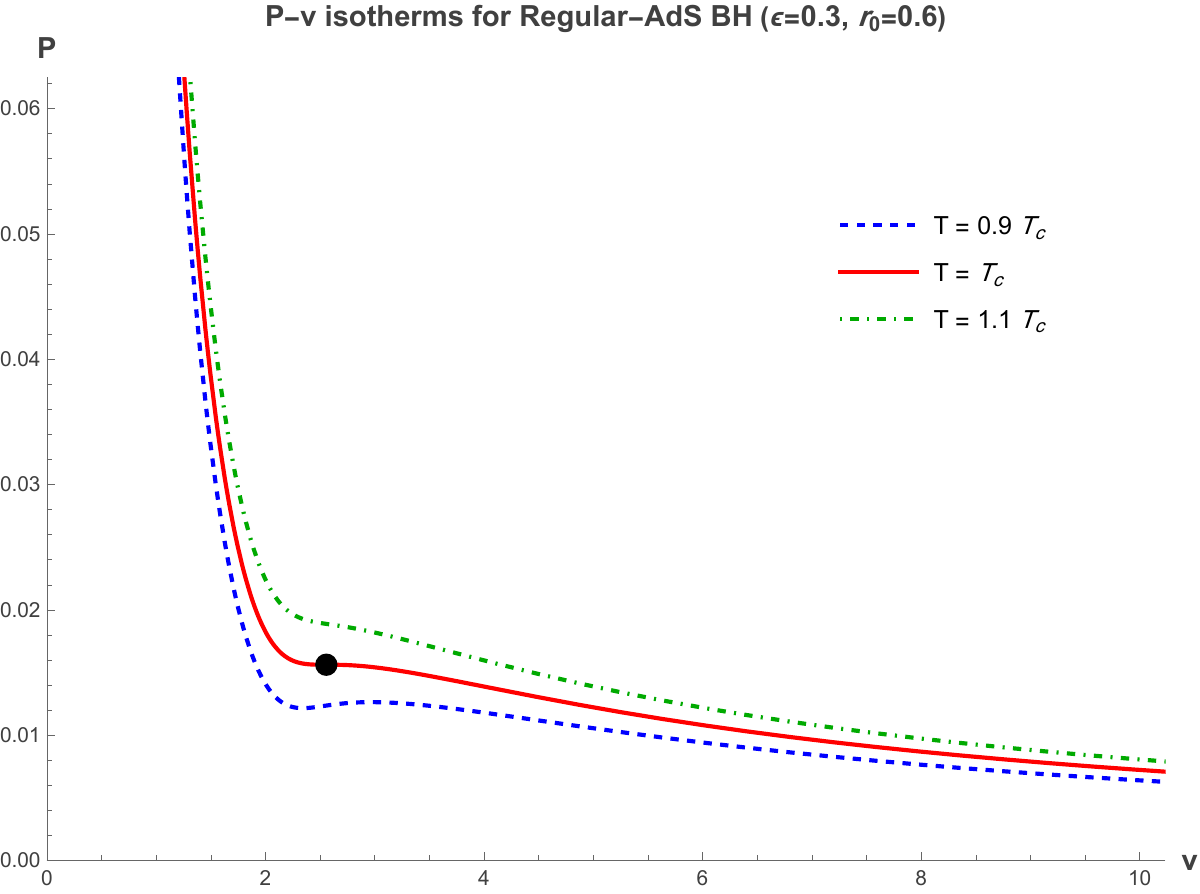}
        \caption{$r_0=0.6$}
    \end{subfigure}
    \\ \vspace{0.3cm}
    \begin{subfigure}{0.48\textwidth}
        \includegraphics[width=\textwidth]{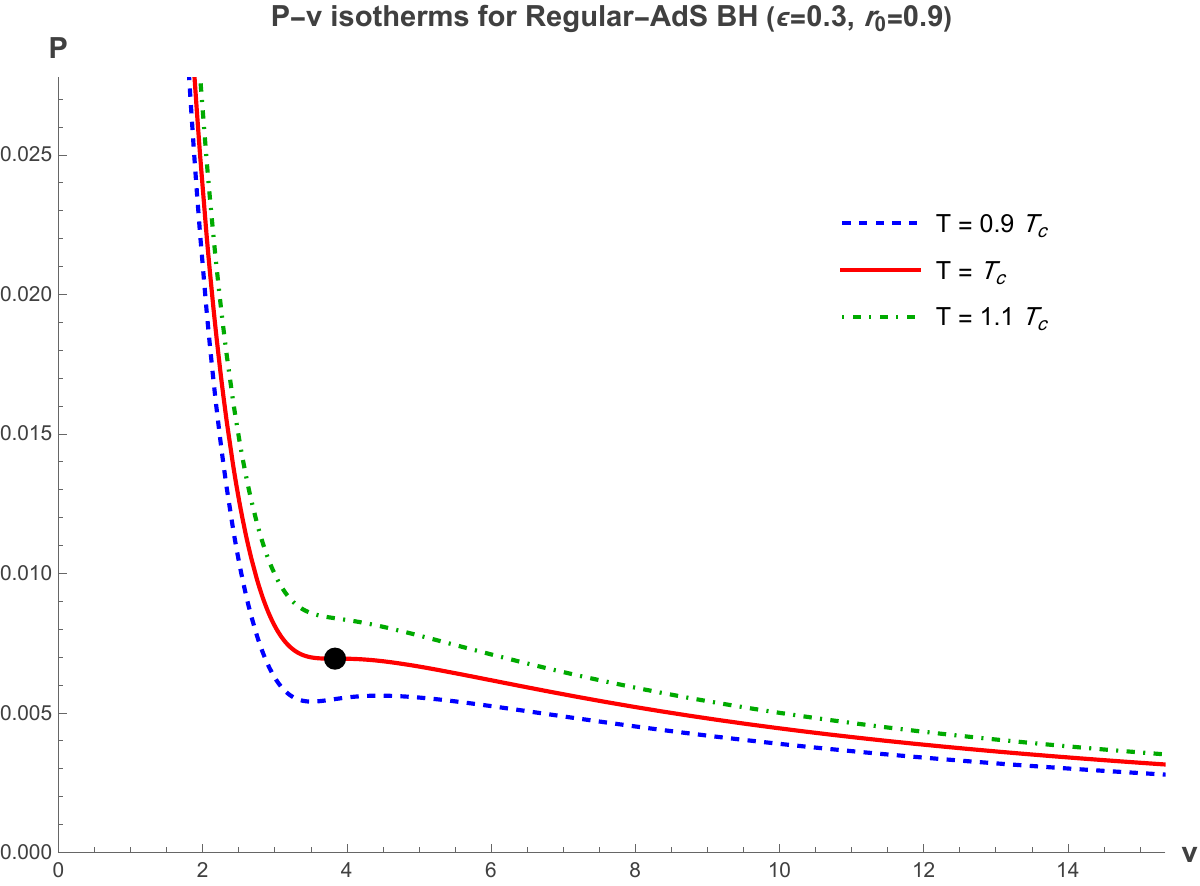}
        \caption{$r_0=0.9$}
    \end{subfigure}
    \hfill
    \begin{subfigure}{0.48\textwidth}
        \includegraphics[width=\textwidth]{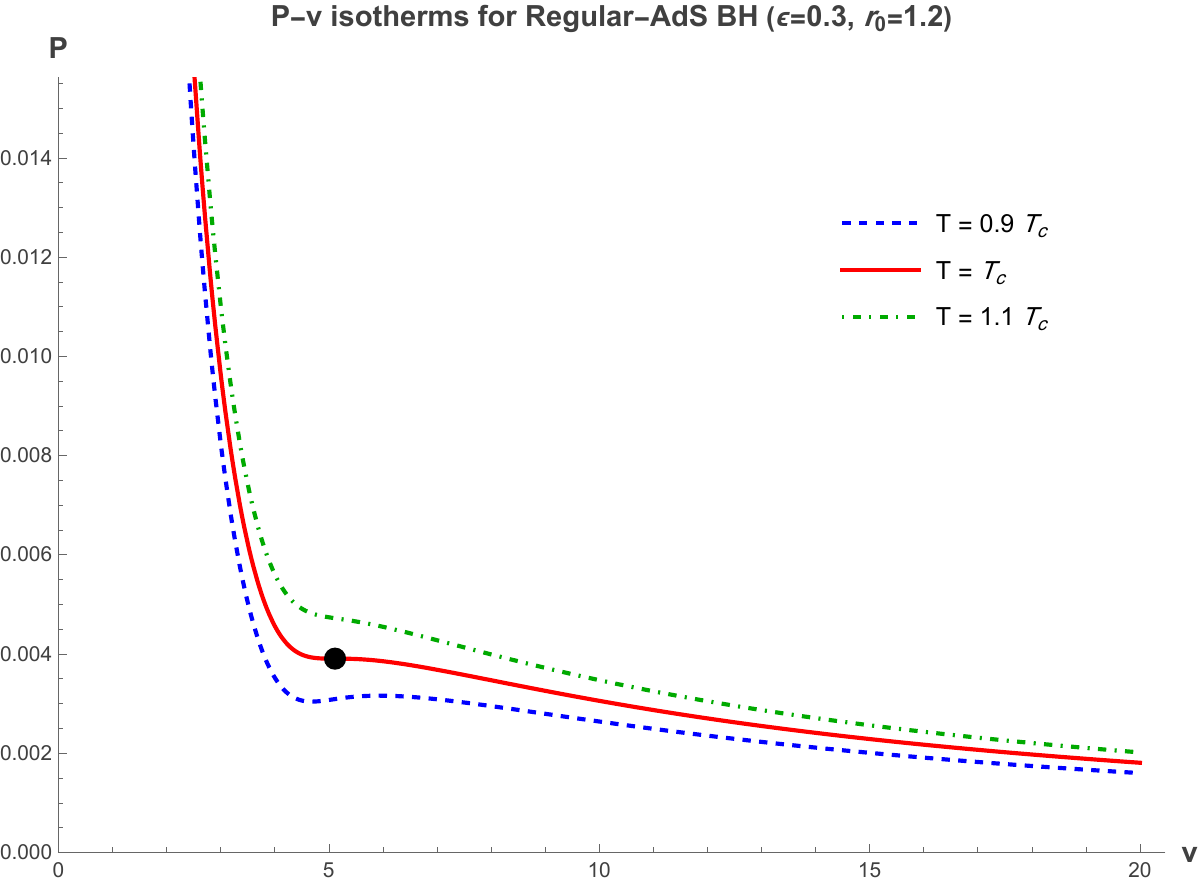}
        \caption{$r_0=1.2$}
    \end{subfigure}
    \caption{The $P-v$ isotherms for a fixed string cloud $\epsilon=0.3$ and varying regularizing core $r_0$.}
    \label{fig:pv_diagrams_r0}
\end{figure}

The characteristics of this equation of state are illustrated in the multi-panel $P-v$ diagrams presented in Figs.~\ref{fig:pv_diagrams_eps} and \ref{fig:pv_diagrams_r0}. At sub-critical temperatures ($T < T_c$), the isotherms display an oscillatory pattern, marked by a local minimum and maximum. This mechanical instability necessitates a Maxwell equal-area construction, thereby validating a first-order phase transition between a Small Black Hole and a Large Black Hole phase \cite{Chamblin1999}. 

\subsection{Critical Behavior and Scale Invariance}

To quantify the impact of the parameters on the phase transitions, we numerically solve for the critical points, satisfying the inflection conditions $\left( \frac{\partial P}{\partial v} \right)_{T} = 0$ and $\left( \frac{\partial^2 P}{\partial v^2} \right)_{T} = 0$. Using the standard proxy $v_c \approx 2r_c$ for the ratio calculation, we first trace the critical coordinates for varying $\epsilon$ while maintaining a fixed core $r_0=0.5$ (Table \ref{tab:critical_points_eps}), and subsequently for varying $r_0$ with fixed $\epsilon=0.3$ (Table \ref{tab:critical_points_r0}).

\begin{table}[h!]
    \centering
    \renewcommand{\arraystretch}{1.3}
    \begin{tabular}{lccccc}
        \toprule
        \textbf{$\epsilon$} & \textbf{$r_0$} & $r_c$ & $T_c$ & $P_c$ & $\rho_c = \frac{P_c (2 r_c)}{T_c}$ \\
        \midrule
        $0$ & $0.5$ & $1.05034$ & $0.144705$ & $0.0329418$ & $0.477796$ \\
        $0.3$ & $0.5$ & $1.06650$ & $0.100036$ & $0.0224871$ & $0.479191$ \\
        $0.6$ & $0.5$ & $1.09200$ & $0.0560438$ & $0.0123488$ & $0.481078$ \\
        $0.9$ & $0.5$ & $1.15234$ & $0.0133729$ & $0.0028114$ & $0.484494$ \\
        \bottomrule
    \end{tabular}
    \caption{Critical point data for the Regular Schwarzschild-AdS BH with CS with fixed  $r_0=0.5$ and varying string cloud parameter $\epsilon$. The critical radius $r_0$, temperature $T_c$, and pressure $P_c$. The compressibility ratio $\rho_c$ . The critical ratio $\rho_c$ modifies dynamically with respect to the string density.}
    \label{tab:critical_points_eps}
\end{table}

\begin{table}[h!]
    \centering
    \renewcommand{\arraystretch}{1.3}
    \begin{tabular}{lccccc}
        \toprule
        \textbf{$\epsilon$} & \textbf{$r_0$} & $r_c$ & $T_c$ & $P_c$ & $\rho_c = \frac{P_c (2 r_c)}{T_c}$ \\
        \midrule
        $0.3$ & $0.4$ & $0.853199$ & $0.125045$ & $0.0351360$ & $0.479191$ \\
        $0.3$ & $0.6$ & $1.279800$ & $0.0833635$ & $0.0156160$ & $0.479191$ \\
        $0.3$ & $0.9$ & $1.919700$ & $0.0555757$ & $0.0069404$ & $0.479191$ \\
        $0.3$ & $1.2$ & $2.559600$ & $0.0416817$ & $0.0039040$ & $0.479191$ \\
        \bottomrule
    \end{tabular}
    \caption{ Critical point data for the Regular Schwarzschild-AdS BH with CS with fixed $\epsilon=0.3$ and varying the regularization parameter $r_0$. The critical radius $r_0$, temperature $T_c$, and pressure $P_c$. The compressibility ratio $\rho_c$ . The critical ratio $\rho_c$ remains strictly scale-invariant.}
    \label{tab:critical_points_r0}
\end{table}

 As the string density $\epsilon$ increases, $r_c$ shifts outwards, while $T_c$ and $P_c$ drop. The string cloud exerts a negative radial pressure, making it easier for the fluid to expand, thus inducing phase transitions at much lower thermal energies \cite{Morais2024}.
   Also for the Table \ref{tab:critical_points_r0} While varying $r_0$ drastically alters the values of $r_c$, $T_c$, and $P_c$, the compressibility ratio $\rho_c$ remains strictly constant at $\rho_c = 0.479191$. Since $r_0$ carries dimensions of length, it strictly serves as a scaling factor for the geometry where, by dimensional analysis, $r_c \propto r_0$, $T_c \propto 1/r_0$, and $P_c \propto 1/r_0^2$. Consequently, the ratio $\rho_c \propto (1/r_0^2 \cdot r_0) / (1/r_0) = 1$ perfectly cancels out the length scale \cite{Gunasekaran2012}. 
   In standard Reissner-Nordström-AdS black holes, the ratio perfectly matches the classical Van der Waals value $\rho_c = 3/8 = 0.375$. In our model, the quantum-inspired core stiffness raises the base ratio to $\approx 0.477$. Crucially, because $\epsilon$ is dimensionless, it fundamentally alters the balance of the equation of state rather than just scaling it, driving the ratio up to $\approx 0.484$ \cite{Hendi2017}.

\subsection{Critical Exponents}
\label{subsec:critical_exponents}

To completely classify the phase transition of our regular string-cloud black hole, we calculate the critical exponents $\alpha$, $\beta$, $\gamma$, and $\delta$. These exponents characterize the behavior of physical quantities in the infinitesimal vicinity of the critical point $(T_c, P_c, v_c)$. We begin by defining the dimensionless reduced thermodynamic variables:
\begin{equation}
    t = \frac{T}{T_c} - 1, \quad p = \frac{P}{P_c} - 1, \quad \omega = \frac{v_{\text{eff}}}{v_c} - 1.
\end{equation}
Here, the variables are expanded around the critical point such that $t$, $p$, and $\omega$ are infinitesimally small. By substituting these reduced variables into our equation of state and performing a Taylor series expansion near the critical point ($t \approx 0, \omega \approx 0$), the reduced pressure takes the standard form:
\begin{equation}
    p \approx A t - B t \omega - C \omega^3 + \mathcal{O}(t \omega^2, \omega^4),
\end{equation}
where $A$, $B$, and $C$ are strictly positive constants determined by the critical parameters $T_c$, $P_c$, and $v_c$. 
We can now systematically derive the four critical exponents \cite{Banerjee2012, Majhi2017}, As a consequence, the critical exponents take the mean-field value :
\begin{itemize}
    \item The exponent $\alpha$ dictates the behavior of the specific heat at constant volume, $C_v \propto |t|^{-\alpha}$. Since the entropy $S$ of our regular black hole depends exclusively on the horizon radius , the specific heat at constant volume strictly vanishes ($C_v = 0$). This directly yields $\alpha = 0$.
    \item  The exponent $\beta$ characterizes the behavior of the order parameter, which is the difference in specific volumes between the large and small black hole phases, $\eta = \omega_l - \omega_s \propto |t|^{\beta}$. By applying Maxwell's equal-area law $\int_{\omega_s}^{\omega_l} \omega dp = 0$ to the expanded equation of state and evaluating it at a constant sub-critical pressure ($p = \text{const}$), we obtain $\omega_l \approx -\omega_s \propto \sqrt{-t}$. Therefore, $\beta = 1/2$.
    \item The exponent $\gamma$ governs the isothermal compressibility, $\kappa_T = -\frac{1}{v} \left( \frac{\partial v}{\partial P} \right)_T \propto |t|^{-\gamma}$. By differentiating our expanded equation of state with respect to $\omega$, we find that $\left( \frac{\partial p}{\partial \omega} \right)_t \approx -B t$. The compressibility is proportional to the inverse of this derivative, leading to $\kappa_T \propto 1/t$. Hence, $\gamma = 1$.
    \item The exponent $\delta$ is defined along the critical isotherm ($t = 0$), where $|p| \propto |\omega|^{\delta}$. Setting $t=0$ in our Taylor expansion trivially reduces the equation of state to $p \approx -C \omega^3$. This yields $\delta = 3$.
\end{itemize}

Remarkably, these derived critical exponents ($\alpha=0$, $\beta=1/2$, $\gamma=1$, $\delta=3$) perfectly match the standard values of classical mean-field theory. This proves that while the string cloud background and the regularizing core significantly alter the critical values and the critical compressibility ratio $\rho_c$ \cite{Ma2016}, they do not change the fundamental universality class of the thermodynamic phase transition, which remains firmly within the Van der Waals domain.

\section{Thermodynamic Stability and Phase Transitions}
\label{sec:stability}

The global preferred state of a thermodynamic system at constant pressure and temperature is governed by the minimization of the Gibbs free energy, $G = M - TS$. 

\begin{figure}[h!]
    \centering
    \begin{subfigure}{0.48\textwidth}
        \includegraphics[width=\textwidth]{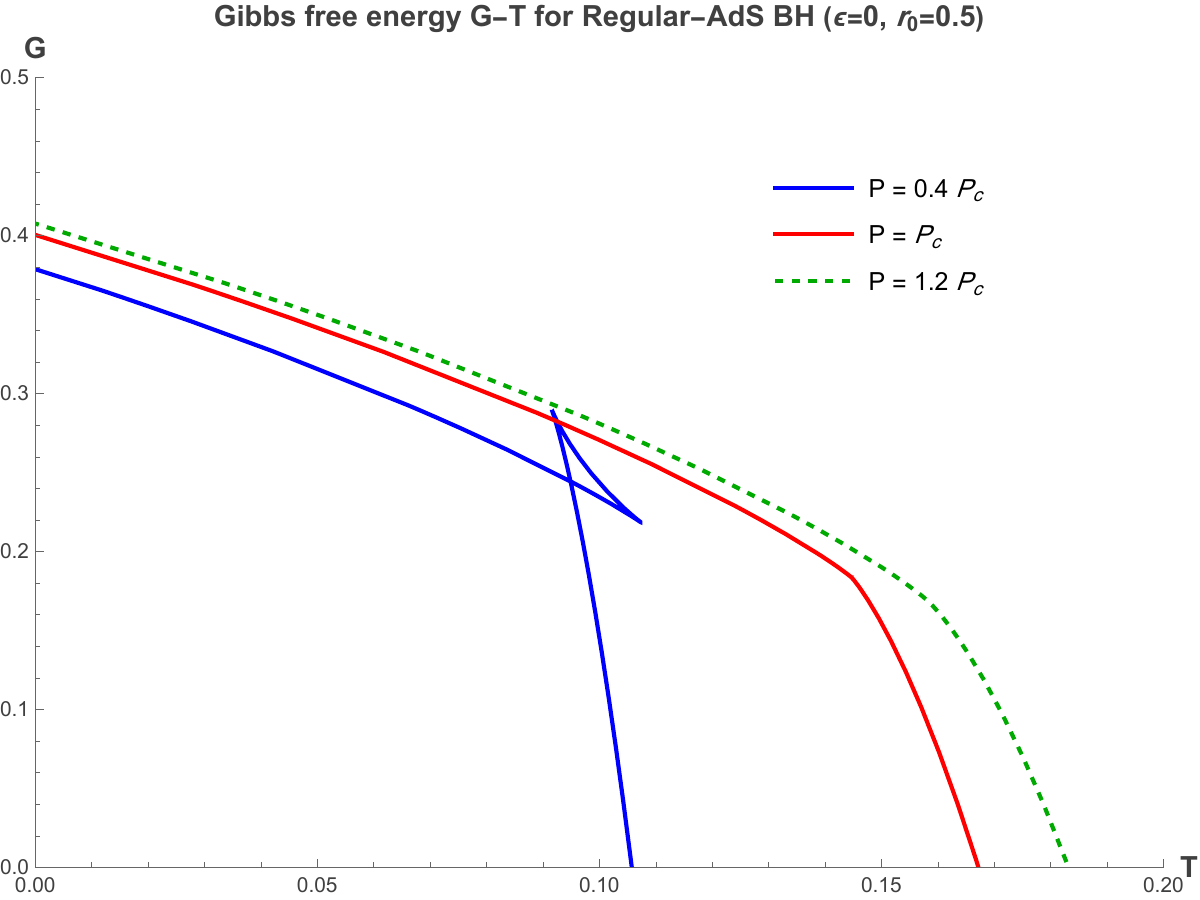}
        \caption{$\epsilon=0$}
    \end{subfigure}
    \hfill
    \begin{subfigure}{0.48\textwidth}
        \includegraphics[width=\textwidth]{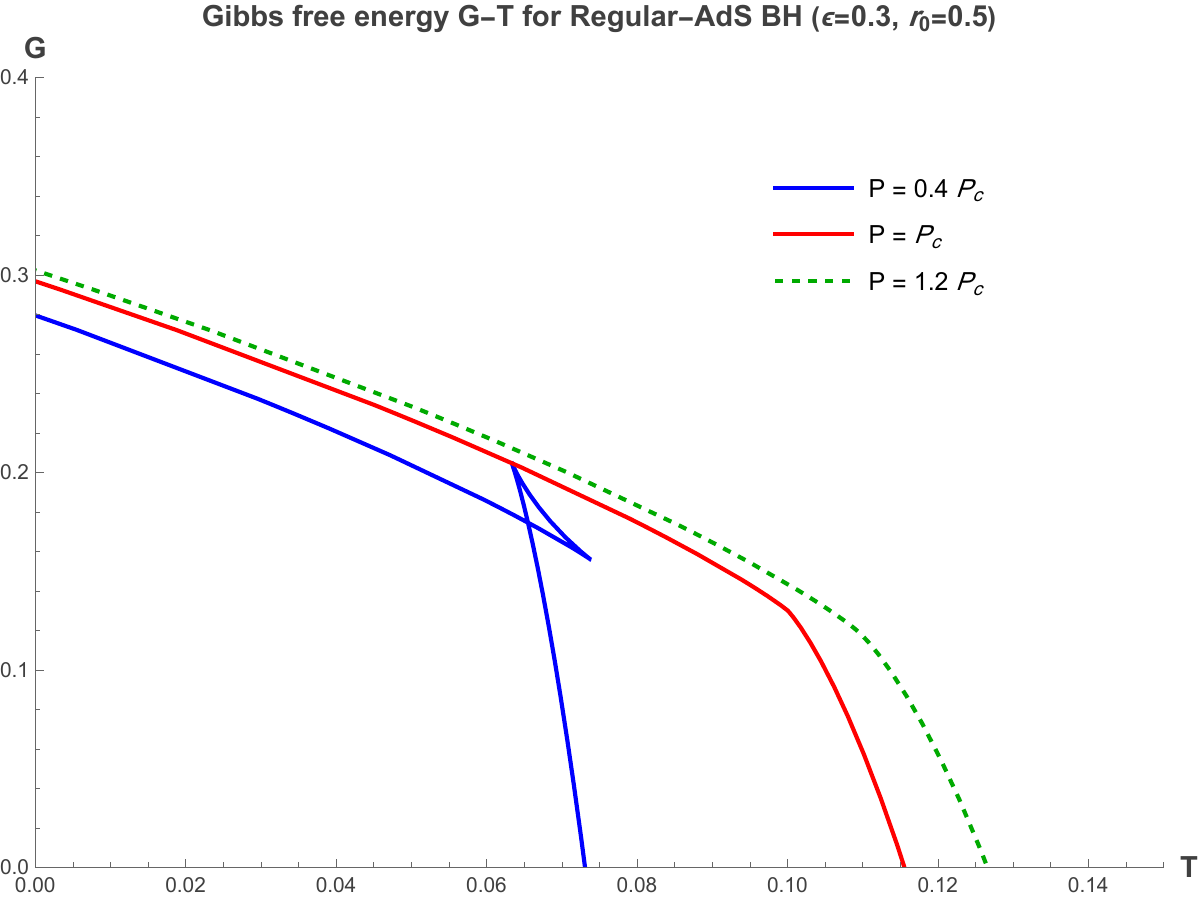}
        \caption{$\epsilon=0.3$}
    \end{subfigure}
    \\ \vspace{0.3cm}
    \begin{subfigure}{0.48\textwidth}
        \includegraphics[width=\textwidth]{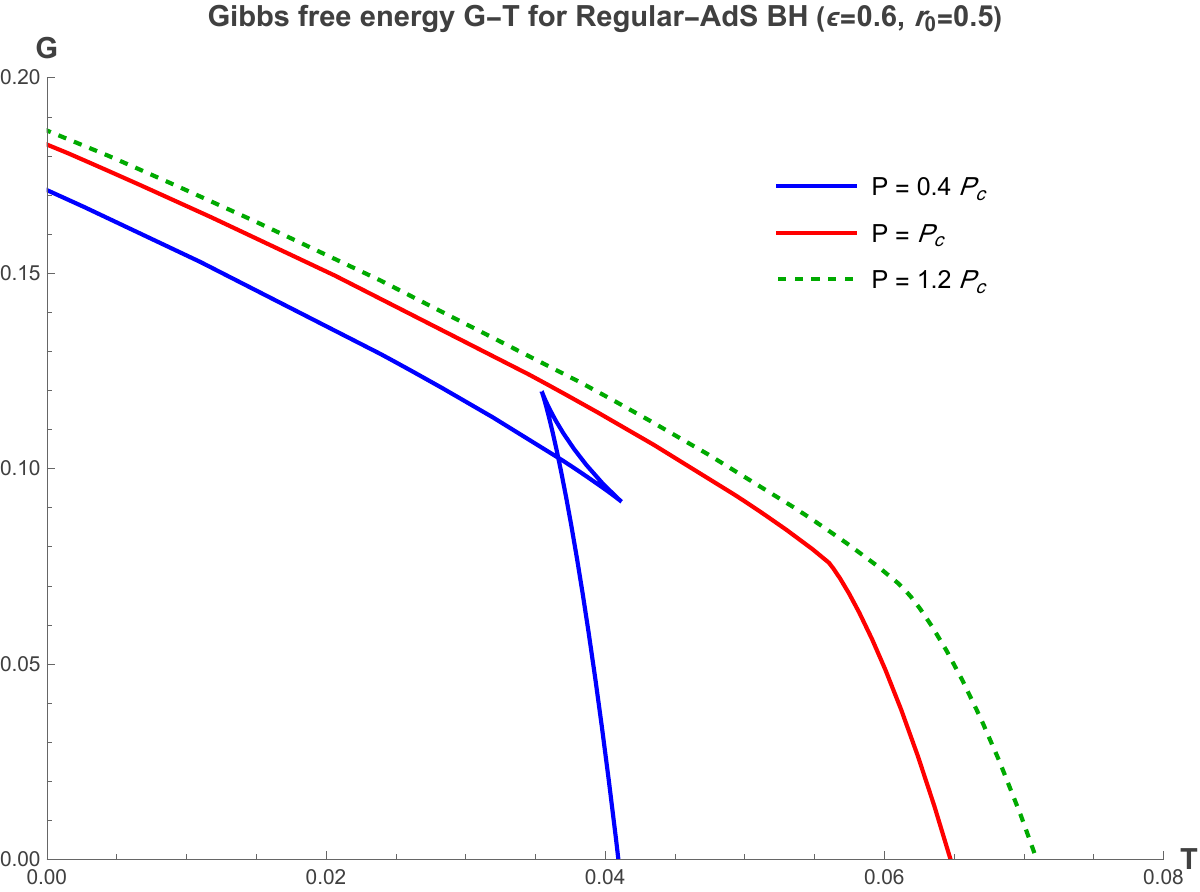}
        \caption{$\epsilon=0.6$}
    \end{subfigure}
    \hfill
    \begin{subfigure}{0.48\textwidth}
        \includegraphics[width=\textwidth]{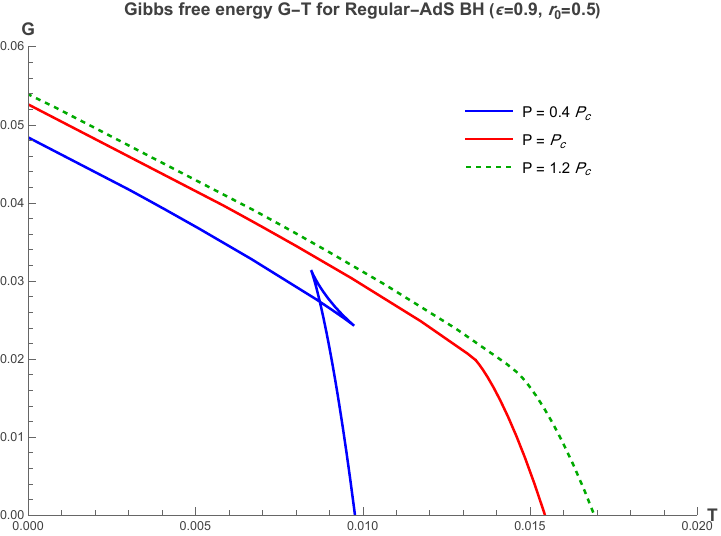}
        \caption{$\epsilon=0.9$}
    \end{subfigure}
    \caption{Gibbs free energy $G$ vs $T$ for varying $\epsilon$. }
    \label{fig:gibbs_diagrams_eps}
\end{figure}

\begin{figure}[h!]
    \centering
    \begin{subfigure}{0.48\textwidth}
        \includegraphics[width=\textwidth]{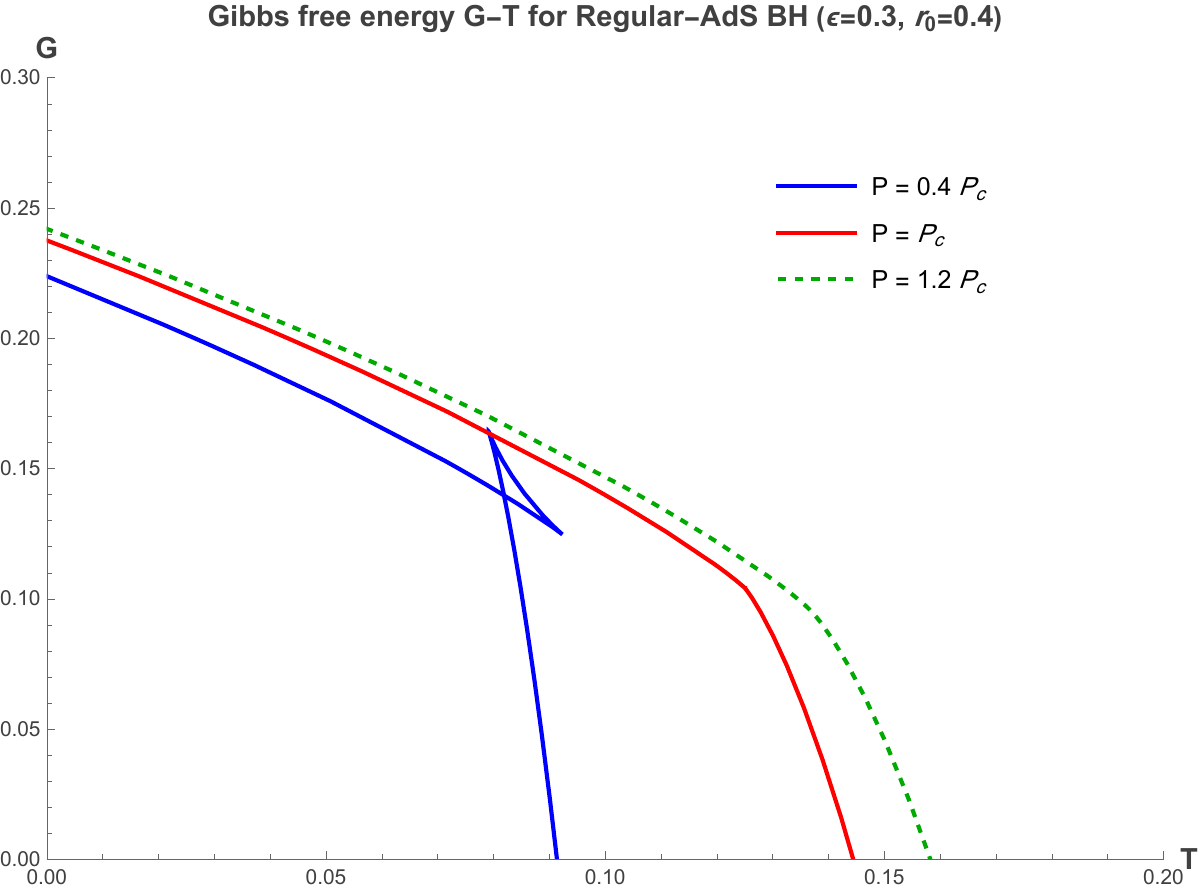}
        \caption{$r_0=0.4$}
    \end{subfigure}
    \hfill
    \begin{subfigure}{0.48\textwidth}
        \includegraphics[width=\textwidth]{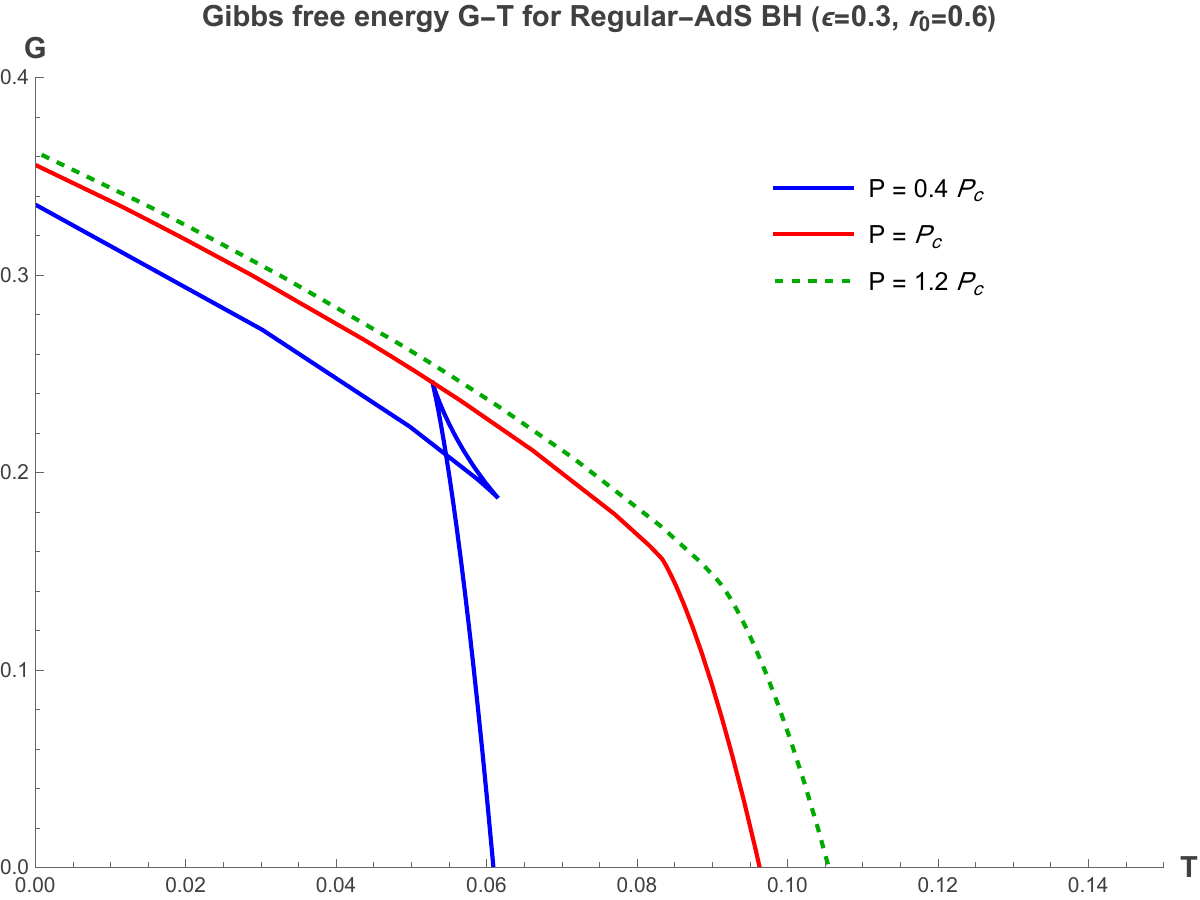}
        \caption{$r_0=0.6$}
    \end{subfigure}
    \\ \vspace{0.3cm}
    \begin{subfigure}{0.48\textwidth}
        \includegraphics[width=\textwidth]{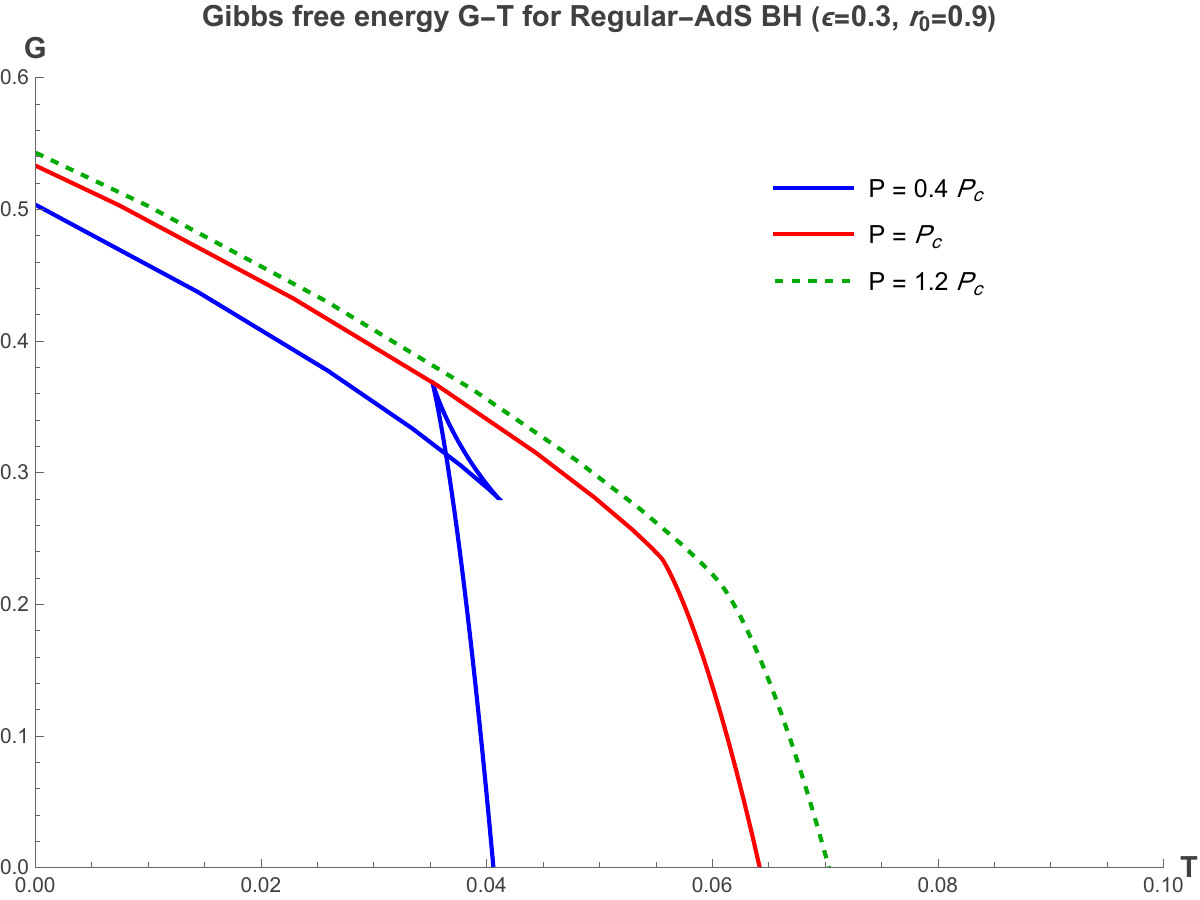}
        \caption{$r_0=0.9$}
    \end{subfigure}
    \hfill
    \begin{subfigure}{0.48\textwidth}
        \includegraphics[width=\textwidth]{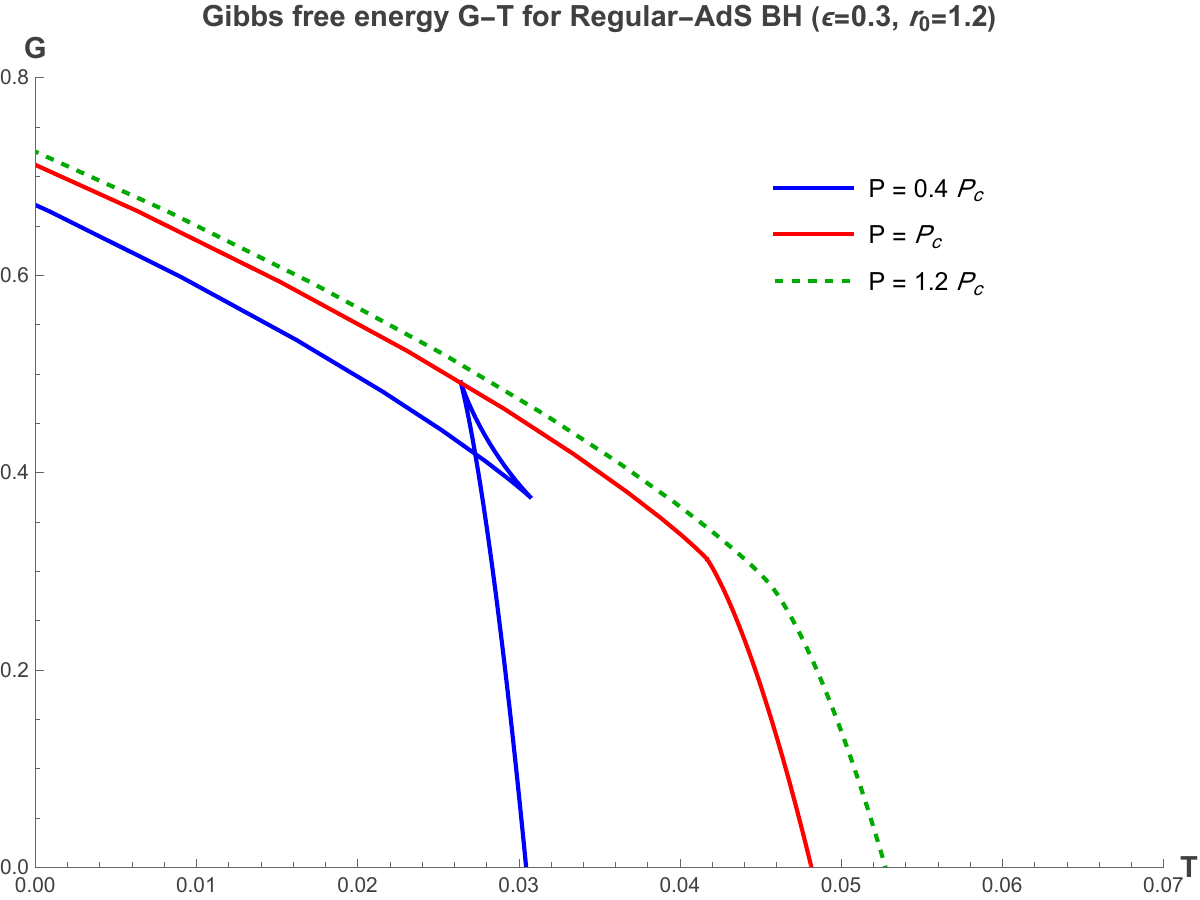}
        \caption{$r_0=1.2$}
    \end{subfigure}
    \caption{Gibbs free energy $G$ vs $T$ for fixed $\epsilon=0.3$ and varying $r_0$. }
    \label{fig:gibbs_diagrams_r0}
\end{figure}

The $G-T$ profiles plotted in Figs.~\ref{fig:gibbs_diagrams_eps} and \ref{fig:gibbs_diagrams_r0} demonstrate that for pressures below the critical value ($P = 0.4 P_c$), the free energy curve develops a multi-valued "swallowtail" catastrophe. This topological feature corresponds to the unstable oscillating region in the $P-v$ diagrams and confirms the transition from the SBH state to the LBH state at the crossing point where the free energies of the two phases are equal \cite{Kubiznak2012}.

 The coexistence line delineates the $(T, P)$ coordinates at which the first-order transition between the SBH and LBH phases transpires.

\begin{figure}[h!]
    \centering
    \includegraphics[width=0.75\textwidth]{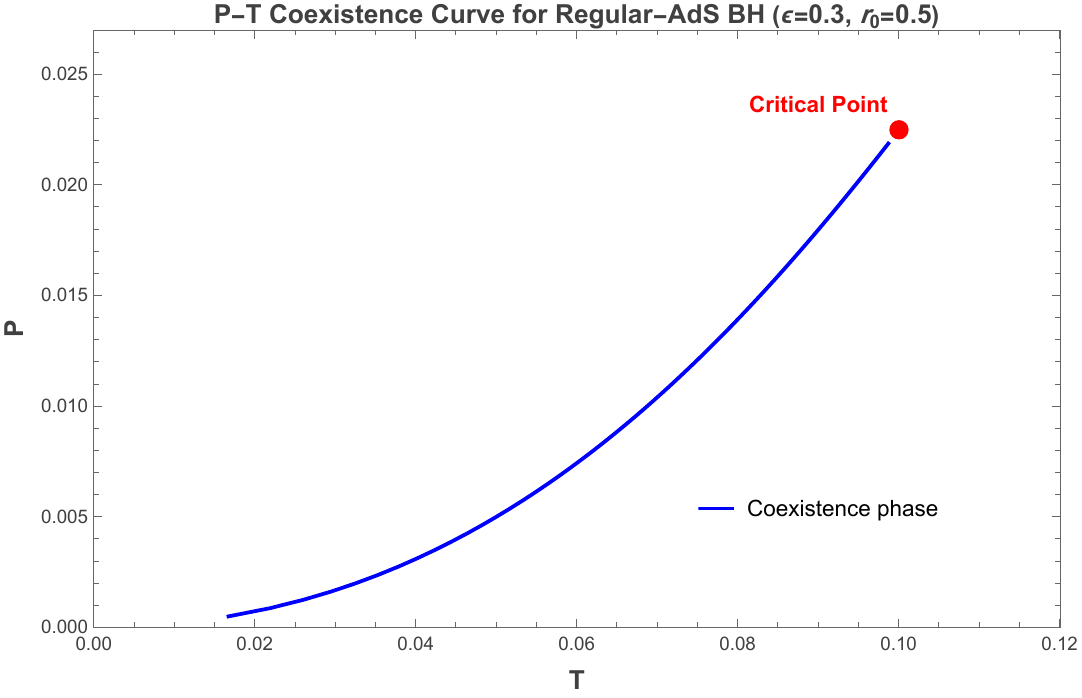}
    \caption{The $P-T$ coexistence curve for the regular AdS black hole with string cloud ($\epsilon=0.3$, $r_0=0.5$). The curve separates the SBH phase (upper left) from the LBH phase (lower right) and terminates strictly at the critical point $(T_c, P_c)$.}
    \label{fig:coexistence}
\end{figure}

The coexistence curve, illustrated in Fig.~\ref{fig:coexistence}, begins at low values and increases monotonically until it abruptly concludes at the crucial point $(T_c, P_c)$. Beyond this threshold, the differentiation between small and large black holes vanishes, transitioning into a supercritical phase. This curve is dictated by the conventional Clapeyron equation \cite{Gunasekaran2012}:
\begin{equation}
    \frac{dP}{dT} = \frac{\Delta S}{\Delta V},
\end{equation}
where $\Delta S = S_{LBH} - S_{SBH}$ and $\Delta V = V_{LBH} - V_{SBH}$. Since both the entropy and volume of the large black hole are greater than those of the small black hole, the slope $\frac{dP}{dT}$ is strictly positive, perfectly matching the physics of the classic liquid-gas phase diagram \cite{Kubiznak2012}.

To ascertain the local thermodynamic stability of the black hole, we evaluate the isobaric heat capacity $C_P$. A thermodynamic system is locally stable against thermal fluctuations if its heat capacity is strictly positive ($C_P > 0$), whereas a negative heat capacity ($C_P < 0$) indicates local instability. The heat capacity at constant pressure is defined as:
\begin{equation}
    C_P = T \left( \frac{\partial S}{\partial T} \right)_P = T \frac{\left( \frac{\partial S}{\partial r_+} \right)_P}{\left( \frac{\partial T}{\partial r_+} \right)_P}.
    \label{eq:cp_def}
\end{equation}

\begin{figure}[h!]
    \centering
    \begin{subfigure}{0.48\textwidth}
        \includegraphics[width=\textwidth]{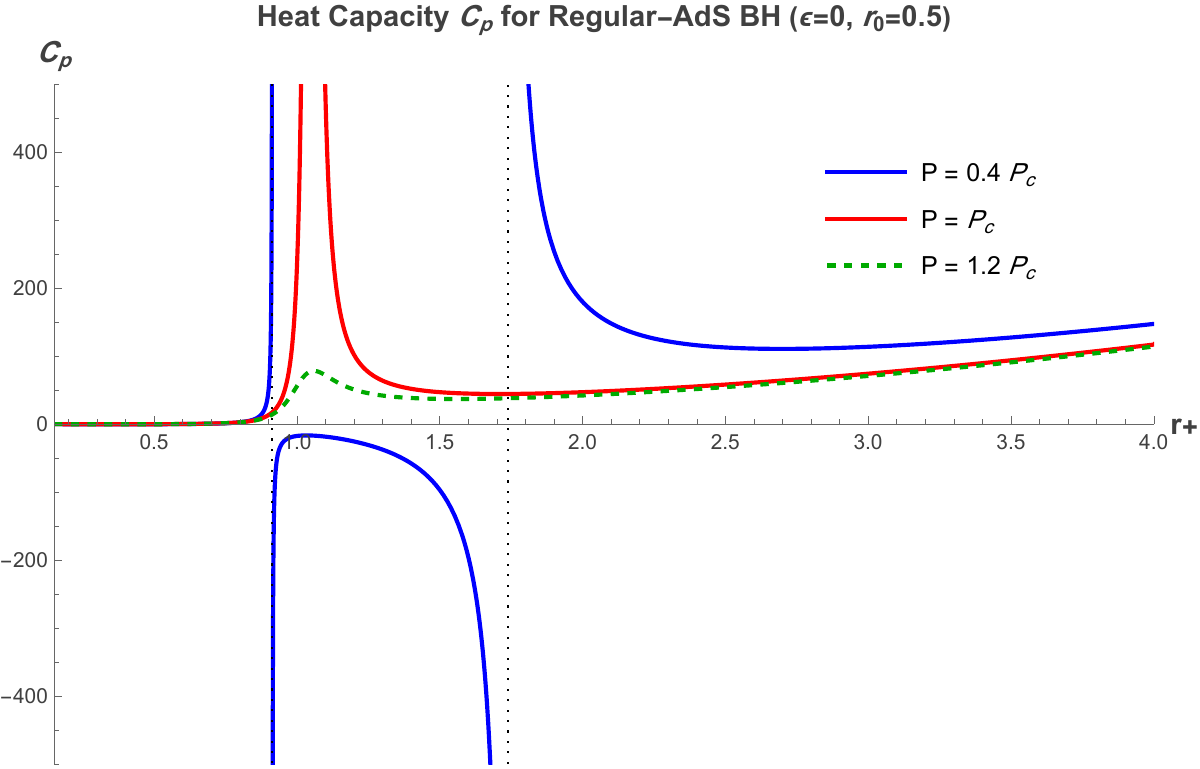}
        \caption{$\epsilon=0$}
    \end{subfigure}
    \hfill
    \begin{subfigure}{0.48\textwidth}
        \includegraphics[width=\textwidth]{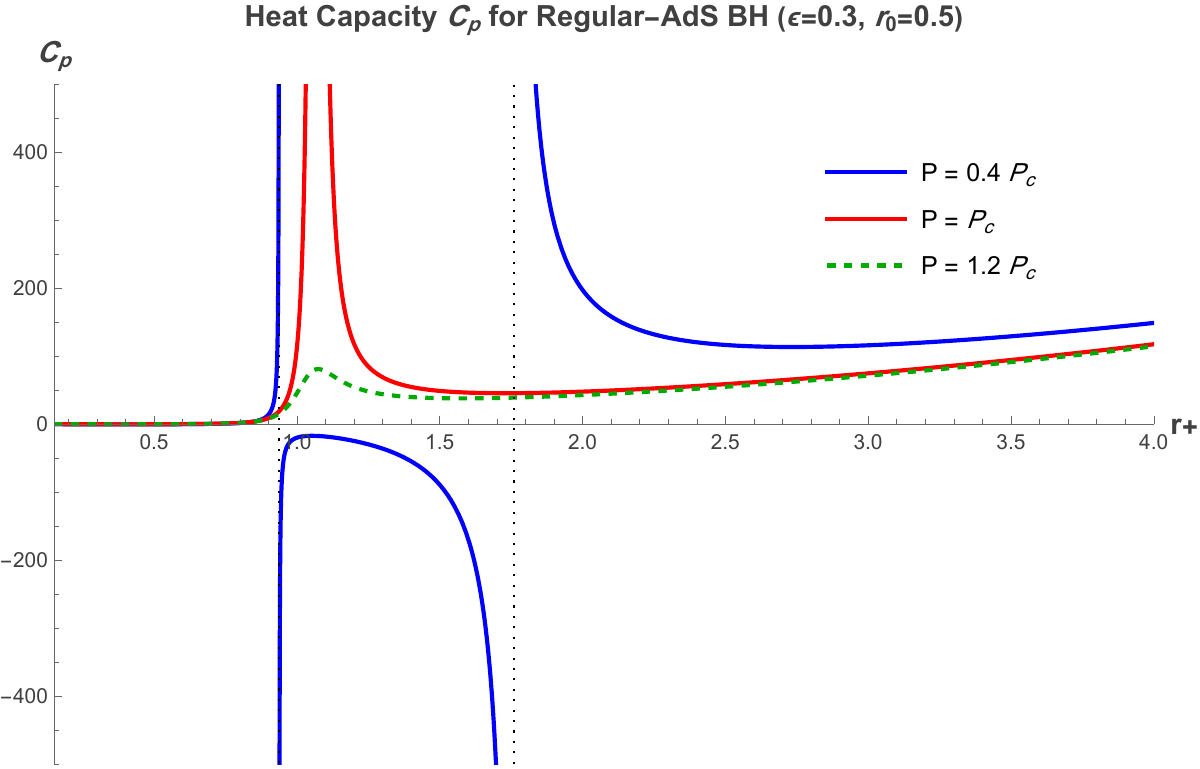}
        \caption{$\epsilon=0.3$}
    \end{subfigure}
    \\ \vspace{0.3cm}
    \begin{subfigure}{0.48\textwidth}
        \includegraphics[width=\textwidth]{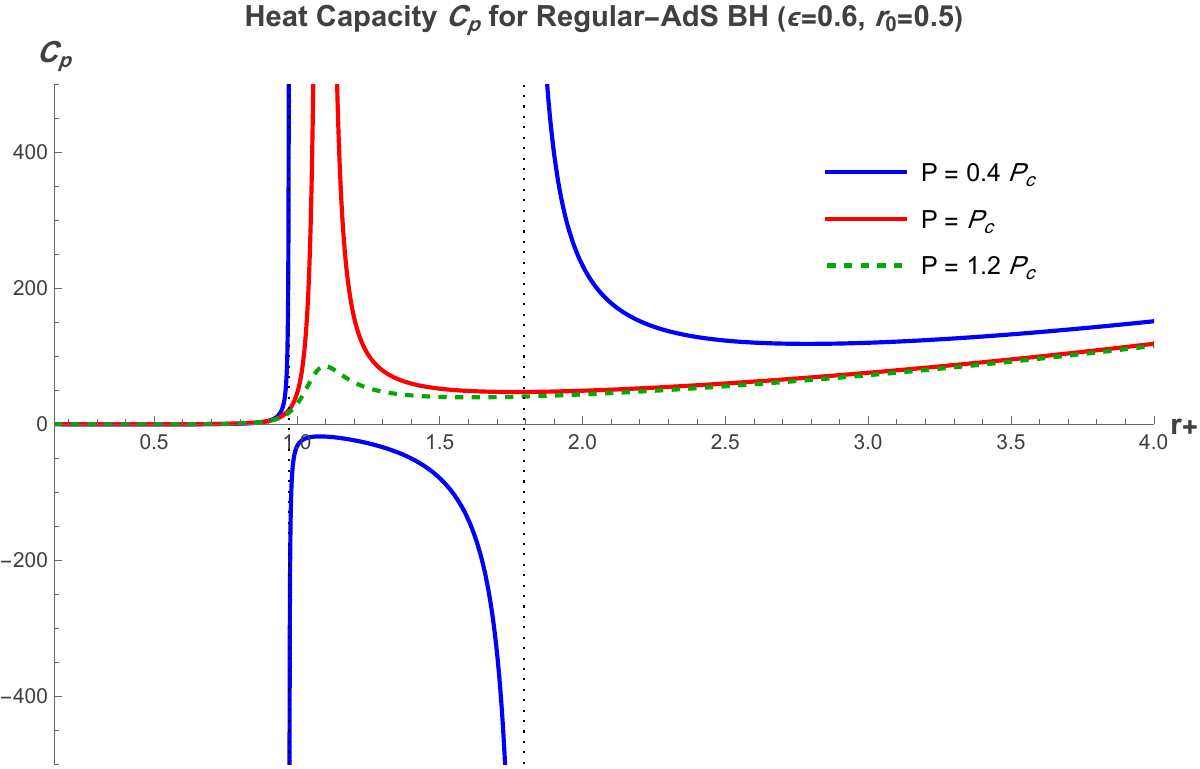}
        \caption{$\epsilon=0.6$}
    \end{subfigure}
    \hfill
    \begin{subfigure}{0.48\textwidth}
        \includegraphics[width=\textwidth]{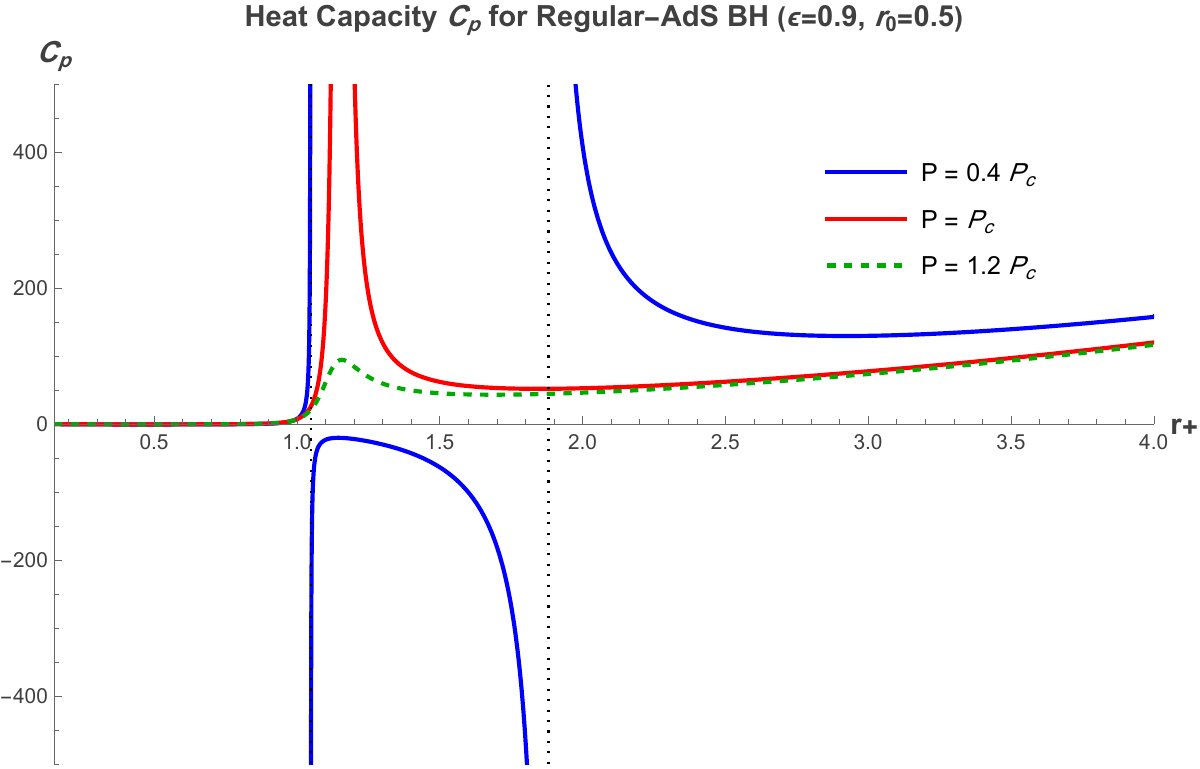}
        \caption{$\epsilon=0.9$}
    \end{subfigure}
    \caption{ Heat capacity $C_P$ versus $r_+$ for varying string cloud parameter $\epsilon$ at fixed $r_0=0.5$.}
    \label{fig:heat_capacity_eps}
\end{figure}

\begin{figure}[h!]
    \centering
    \begin{subfigure}{0.48\textwidth}
        \includegraphics[width=\textwidth]{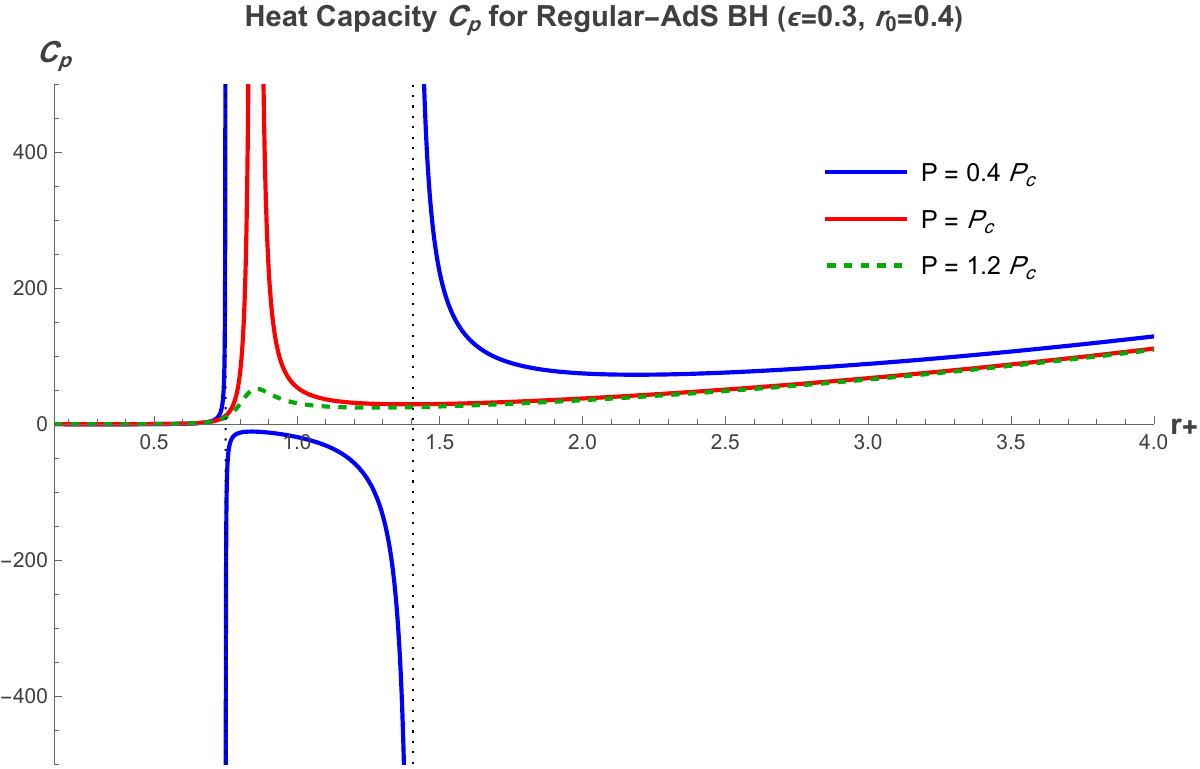}
        \caption{$r_0=0.4$}
    \end{subfigure}
    \hfill
    \begin{subfigure}{0.48\textwidth}
        \includegraphics[width=\textwidth]{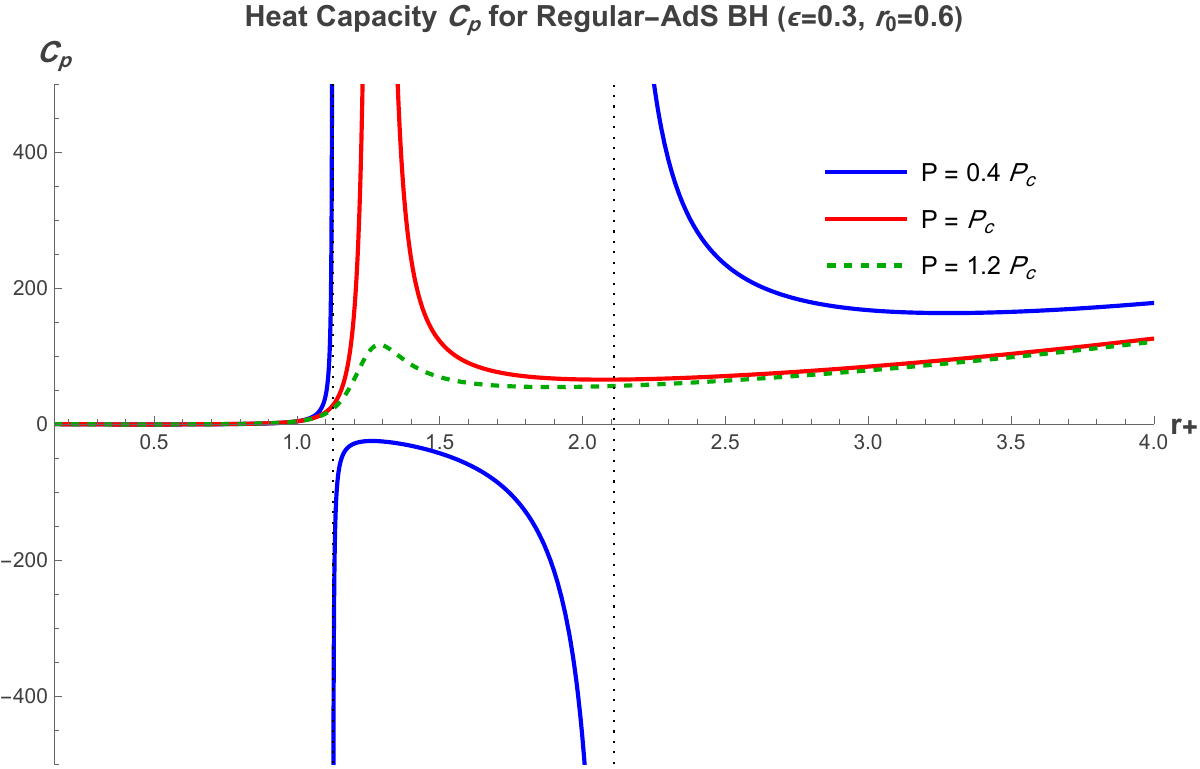}
        \caption{$r_0=0.6$}
    \end{subfigure}
    \\ \vspace{0.3cm}
    \begin{subfigure}{0.48\textwidth}
        \includegraphics[width=\textwidth]{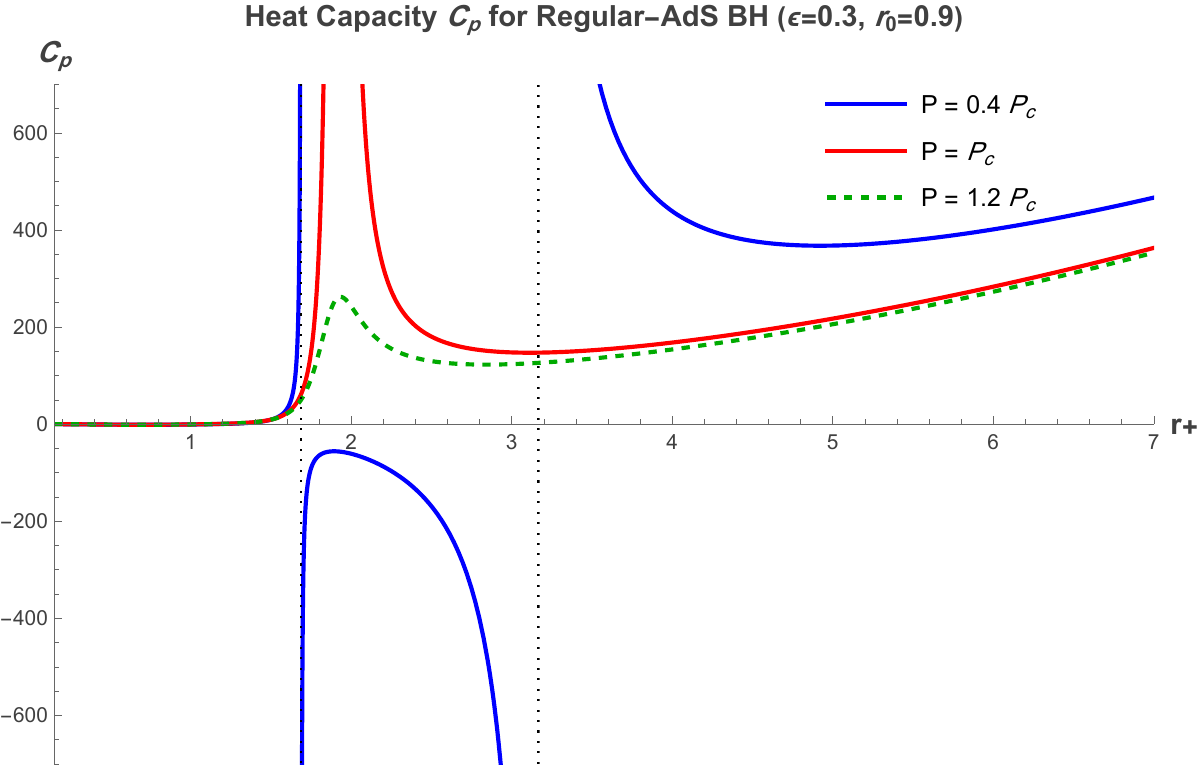}
        \caption{$r_0=0.9$}
    \end{subfigure}
    \hfill
    \begin{subfigure}{0.48\textwidth}
        \includegraphics[width=\textwidth]{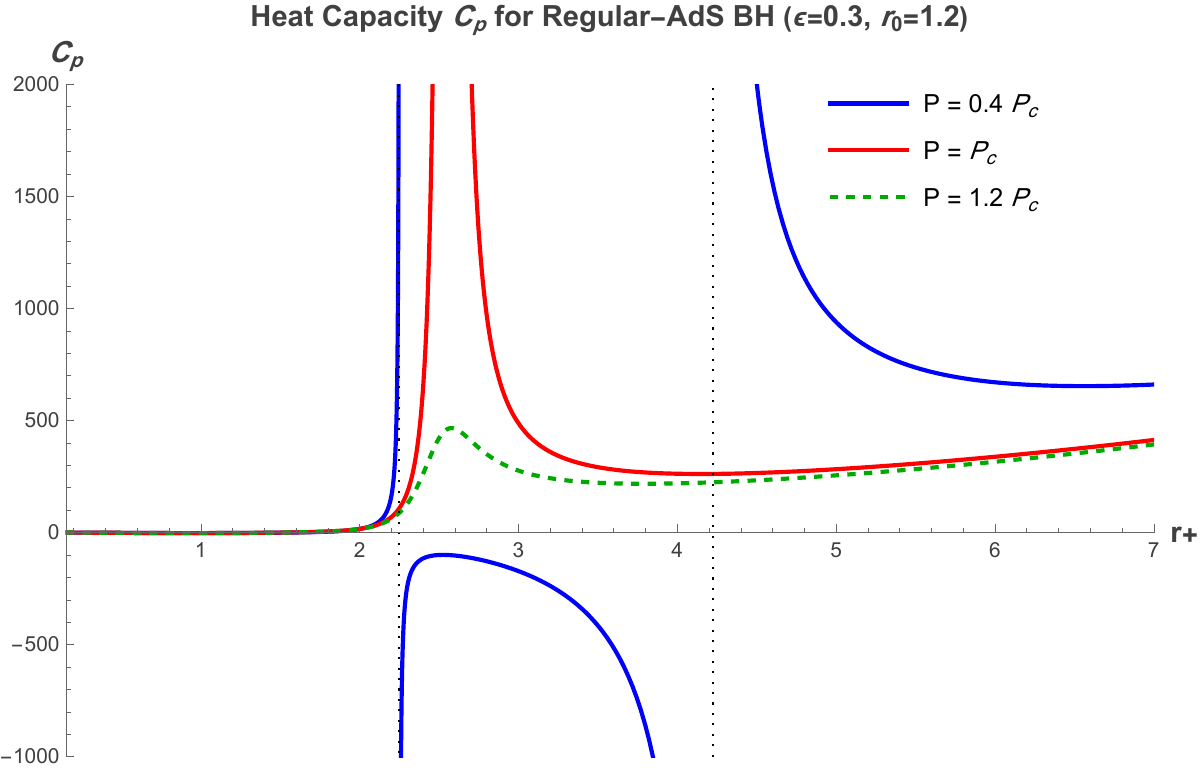}
        \caption{$r_0=1.2$}
    \end{subfigure}
    \caption{Isobaric heat capacity $C_P$ versus $r_+$ for fixed $\epsilon=0.3$ and varying regularizing scale $r_0$.}
    \label{fig:heat_capacity_r0}
\end{figure}

To rigorously evaluate the influence of the string cloud background and the core regularizing length scale, we plot $C_P$ as a function of the horizon radius $r_+$ for varying values of $\epsilon$ (Fig.~\ref{fig:heat_capacity_eps}) and $r_0$ (Fig.~\ref{fig:heat_capacity_r0}). 

The system exhibits classic van der Waals-like thermodynamic behavior across all tested parameters. For sub-critical pressures ($P = 0.4 P_c$), $C_P$ suffers from two infinite discontinuities that partition the phase space into a locally stable SBH branch ($C_P > 0$), a locally unstable intermediate black hole branch ($C_P < 0$), and a locally stable LBH branch ($C_P > 0$). At the critical pressure ($P = P_c$), these divergences coalesce into a single singular point, signaling a second-order phase transition \cite{Spallucci2013}. For super-critical pressures ($P = 1.2 P_c$), the heat capacity remains strictly positive and continuous, reflecting absolute local stability. 

Crucially, the introduction of the string cloud parameter $\epsilon$ systematically modulates the spatial scales of these phase transitions. As $\epsilon$ is incrementally increased from $0$ to $0.9$ (Fig.~\ref{fig:heat_capacity_eps}), the divergent points bounding the unstable intermediate branch shift progressively toward larger values of $r_+$. Physically, this indicates that the background cloud of strings alters the effective energy-momentum distribution, enhancing the repulsive gravitational effects near the core. Consequently, a denser string cloud delays the onset of the stable macroscopic LBH configuration, requiring a larger characteristic horizon size to achieve ultimate thermodynamic stability \cite{Letelier1979, Ma2016, Santos2022}.

Similarly, the structural properties of the phase transition are highly sensitive to the regularization parameter $r_0$ (Fig.~\ref{fig:heat_capacity_r0}). As $r_0$ increases from $0.4$ to $1.2$, the unstable intermediate domain broadens significantly, and the critical merging point shifts to larger horizon radii. Since $r_0$ governs the effective length scale of the quantum-corrected, de Sitter-like core, a larger value propagates the core's repulsive anti-gravity effect further outward into the spacetime. Therefore, the black hole must encapsulate a substantially larger thermodynamic volume to overcome this internal repulsion, deferring the SBH-to-LBH phase transition to macroscopic scales \cite{Hayward2006, Singh:2022xgi, Rehan:2024dsg}.
\newpage
\section{Joule-Thomson Expansion}
\label{sec:jt_expansion}

In classical thermodynamics, the Joule-Thomson (JT) expansion describes the temperature change of a real gas as it is forced through a porous plug or a valve while kept insulated, maintaining constant enthalpy ($H = \text{constant}$). In the extended phase space formalism for AdS black holes, the mass $M$ represents the enthalpy. Therefore, the JT expansion characterizes an isenthalpic expansion of the black hole ($dM = 0$) where the thermodynamic pressure $P$ dynamically decreases \cite{Okcu2017}.

To determine whether the black hole fluid undergoes heating or cooling during this expansion, we evaluate the Joule-Thomson coefficient $\mu_{JT}$, defined mathematically as the isenthalpic derivative of temperature with respect to pressure:
\begin{equation}
    \mu_{JT} = \left( \frac{\partial T}{\partial P} \right)_M = \frac{1}{C_P} \left[ T \left( \frac{\partial V}{\partial T} \right)_P - V \right].
    \label{eq:jt_coefficient}
\end{equation}
During the expansion, the pressure continuously drops ($dP < 0$). Consequently:
\begin{itemize}
    \item If $\mu_{JT} > 0$, the temperature decreases ($dT < 0$), and the black hole is in a \textbf{cooling regime}.
    \item If $\mu_{JT} < 0$, the temperature increases ($dT > 0$), and the black hole is in a \textbf{heating regime}.
\end{itemize}
The boundary separating these two distinct physical regimes is marked by the inversion curve, defined strictly by the condition $\mu_{JT} = 0$, which yields the inversion temperature $T_i$:
\begin{equation}
    T_i = V \left( \frac{\partial T}{\partial V} \right)_P.
\end{equation}
By setting $M$ to be constant, we can utilize Eq.~\eqref{eq:mass_psi} and Eq.~\eqref{eq:temperature_psi} to trace the isenthalpic $T-P$ curves and locate their maxima, which identically define the inversion points.

\begin{figure}[h!]
    \centering
    \includegraphics[width=0.75\textwidth]{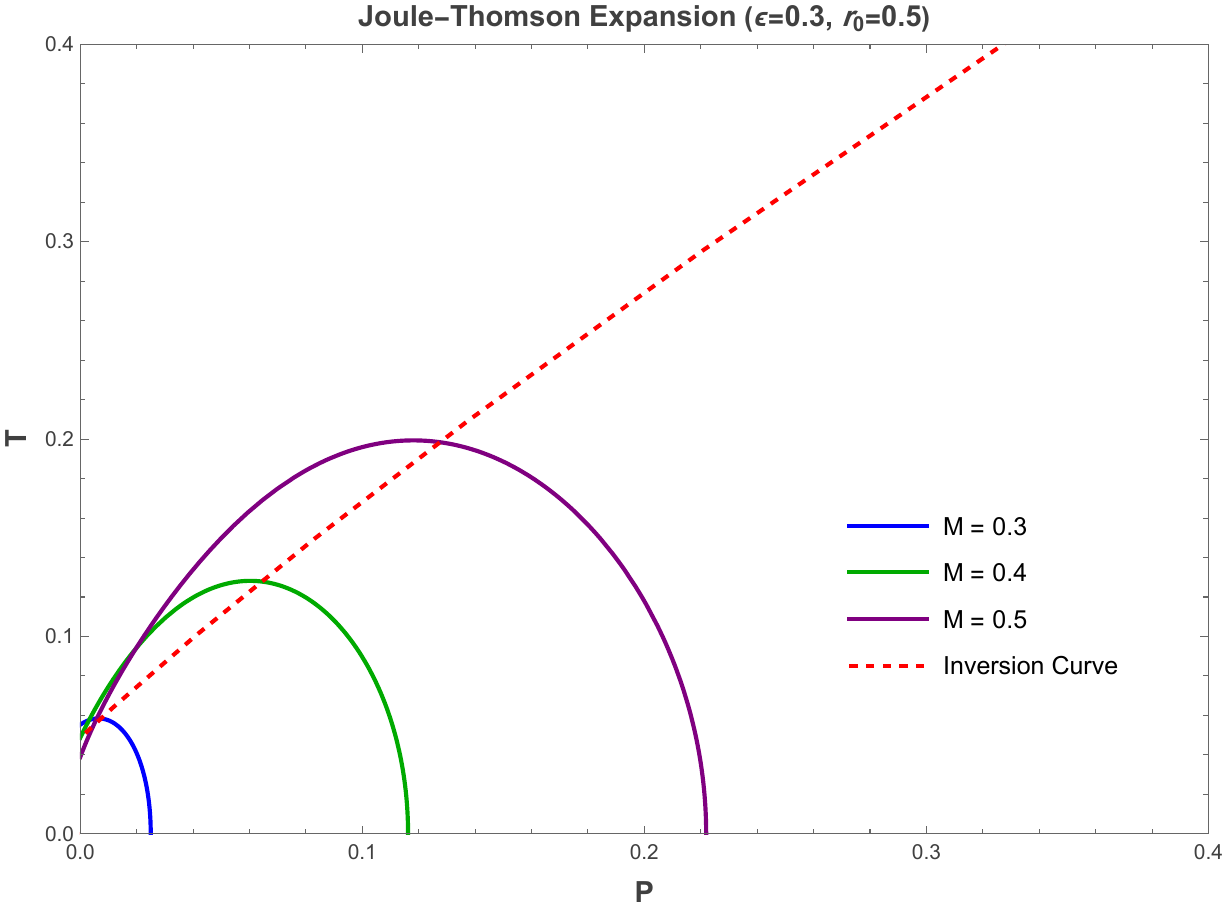}
    \caption{Joule-Thomson expansion ($T-P$ diagram) for the regular AdS black hole with string cloud ($\epsilon=0.3$, $r_0=0.5$). The solid curves represent isenthalpic processes for constant mass $M$. The dashed line represents the inversion curve connecting the maxima of the isenthalpics, separating the cooling region ($\mu_{JT} > 0$, below the curve) from the heating region ($\mu_{JT} < 0$, above the curve).}
    \label{fig:jt_isenthalpic}
\end{figure}

Figure \ref{fig:jt_isenthalpic} maps the isenthalpic curves in the $T-P$ plane for varying constant masses $M = 0.3, 0.4, 0.5$. The inversion curve flawlessly intersects the peak of each isenthalpic curve. The area beneath the inversion curve demonstrates a positive gradient ($\mu_{JT} > 0$), indicating that the black hole cools as it expands (pressure diminishes). In contrast, the area above the inversion curve has a negative slope ($\mu_{JT} < 0$), wherein expansion results in heating \cite{Mo2018}.

\begin{figure}[h!]
    \centering
    \begin{subfigure}{0.48\textwidth}
        \includegraphics[width=\textwidth]{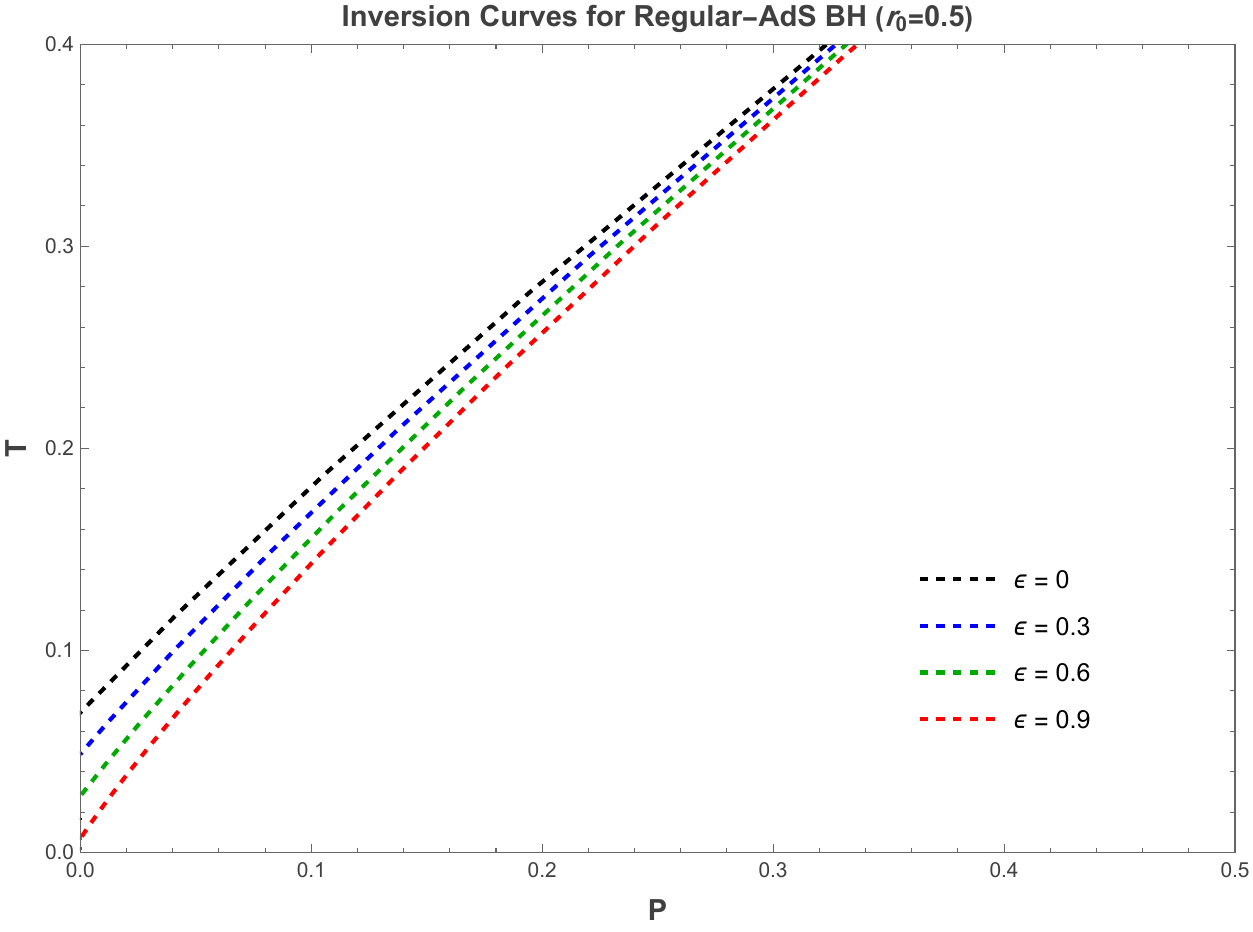}
        \caption{Varying string density $\epsilon$ (fixed $r_0=0.5$)}
    \end{subfigure}
    \hfill
    \begin{subfigure}{0.48\textwidth}
        \includegraphics[width=\textwidth]{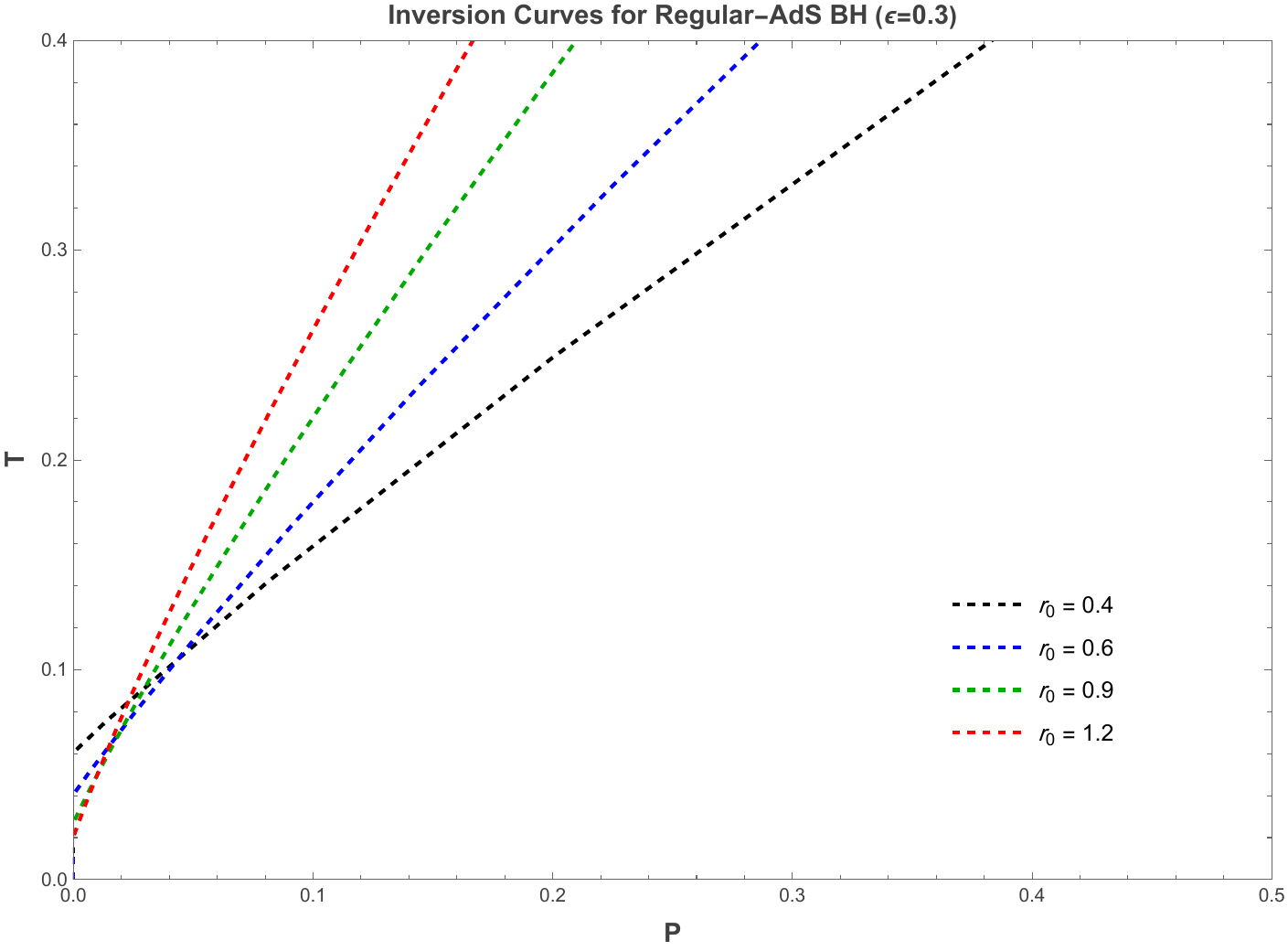}
        \caption{Varying core parameter $r_0$ (fixed $\epsilon=0.3$)}
    \end{subfigure}
    \caption{The inversion curves for the Joule-Thomson expansion. Panel (a) illustrates the influence of the string cloud $\epsilon$, whereas Panel (b) emphasizes the scale effect of the regularization parameter $r_0$.}
    \label{fig:jt_inversion_curves}
\end{figure}
To fully comprehend the influence of backdrop geometry on this process, we illustrated the inversion curves for different string cloud densities $\epsilon$ and regularizing cores $r_0$ in Fig.~\ref{fig:jt_inversion_curves}. 
Panel (a) illustrates that augmenting the string density $\epsilon$ systematically diminishes the inversion curve, markedly reducing the inversion temperature. This demonstrates that the string cloud significantly augments the cooling region of the black hole, complicating the system's transition into a heating regime. In Panel (b), it is evident that as the regularization parameter $r_0$ rises, the inversion curves contract linearly towards the origin. This corroborates our earlier discovery of scale invariance: $r_0$ establishes the overall thermodynamic size of the black hole without fundamentally altering its intrinsic structural ratios \cite{Yerra2018}.

\newpage
\section{Conclusion}
\label{sec:conclusion}

In this paper, we have investigated the extended phase space thermodynamics of a static, spherically symmetric regular black hole surrounded by a string cloud background. By identifying the cosmological constant as a thermodynamic pressure, our analysis yields several pivotal physical conclusions. First, the introduction of the regularizing core $r_0$ fundamentally modifies the horizon structure, generating a Cauchy horizon and preventing the central singularity. Consequently, the black hole possesses a finite, non-zero remnant thermodynamic volume even as the horizon vanishes. Second, we formulated the equation of state by introducing an effective specific volume $v_{\text{}}$. This geometric rescaling allowed us to identify a robust liquid-gas-like first-order phase transition between Small and Large Black Hole states, confirmed globally by the "swallowtail" structure of the Gibbs free energy and locally by the divergences in the isobaric heat capacity. 

A central novelty of our work is the decoupling of the parameter effects on the critical behavior. While the regularization parameter $r_0$ governs the absolute scale of the critical points, it preserves a strict scale invariance, maintaining a constant compressibility ratio $\rho_c \approx 0.479$. In stark contrast, the dimensionless string cloud parameter $\epsilon$ fundamentally alters the nature of the fluid, stretching the horizon and driving the critical ratio upwards. Remarkably, despite these profound geometrical deviations from classical Reissner-Nordström-AdS black holes, we rigorously proved that the critical exponents ($\alpha=0$, $\beta=1/2$, $\gamma=1$, $\delta=3$) remain locked to the classical Van der Waals mean-field universality class. Finally, our mapping of the Joule-Thomson expansion showed that the presence of the string cloud suppresses the inversion curve, thereby amplifying the black hole's cooling regime during isenthalpic expansion.

These findings open several promising avenues for future research. A natural progression would be to apply this extended phase space formalism to the rotating counterpart of this metric, which we recently developed in \cite{Elaima2026}, to assess how spin and the string cloud synergistically affect critical phenomena and Joule-Thomson expansion. Additionally, exploring the holographic dual of these phase transitions via the AdS/CFT correspondence specifically by calculating the holographic entanglement entropy or the two-point correlation functions could provide deeper insights into the underlying quantum microscopic structure of the regularizing core.



\begin{thebibliography}{99}

\bibitem{Bekenstein1973} 
J. D. Bekenstein, \textit{Black holes and entropy}, Phys. Rev. D \textbf{7}, 2333 (1973).

\bibitem{Hawking1975} 
S. W. Hawking, \textit{Particle creation by black holes}, Commun. Math. Phys. \textbf{43}, 199 (1975).

\bibitem{Chamblin1999}
A. Chamblin, R. Emparan, C. V. Johnson, and R. C. Myers, \textit{Holography, thermodynamics, and fluctuations of charged AdS black holes}, Phys. Rev. D \textbf{60}, 104026 (1999).

\bibitem{Kastor2009}
D. Kastor, S. Ray, and J. Traschen, \textit{Enthalpy and the mechanics of AdS black holes}, Class. Quantum Grav. \textbf{26}, 195011 (2009).

\bibitem{Zaslavskii2010}
O.~B.~Zaslavskii,
\textit{Regular black holes and energy conditions,}
Phys. Lett. B \textbf{688}, 278-280 (2010).
[arXiv:1004.2362 [gr-qc]].

\bibitem{Dolan2011}
B. P. Dolan, \textit{The cosmological constant and the black hole equation of state}, Class. Quantum Grav. \textbf{28}, 235017 (2011).

\bibitem{Cvetic2011}
M. Cveti\v{c}, G. W. Gibbons, D. Kubiz\v{n}\'{a}k, and C. N. Pope, \textit{Black hole enthalpy and an entropy inequality for the thermodynamic volume}, Phys. Rev. D \textbf{84}, 024037 (2011).

\bibitem{Kubiznak2012}
D. Kubiz\v{n}\'{a}k and R. B. Mann, \textit{P-V criticality of charged AdS black holes}, JHEP \textbf{2012}, 33 (2012).

\bibitem{Wei2015}
S.-W. Wei and Y.-X. Liu, \textit{Insight into the Microscopic Structure of an AdS Black Hole from a Thermodynamical Phase Transition}, Phys. Rev. Lett. \textbf{115}, 111302 (2015).

\bibitem{Gunasekaran2012}
S. Gunasekaran, R. B. Mann, and D. Kubiz\v{n}\'{a}k, \textit{Extended phase space thermodynamics for charged and rotating black holes and Born-Infeld vacuum polarization}, JHEP \textbf{2012}, 110 (2012).

\bibitem{Altamirano2014}
N. Altamirano, D. Kubiz\v{n}\'{a}k, R. B. Mann, and Z. Sherkatghanad, \textit{Thermodynamics of rotating black holes and black rings: phase transitions and thermodynamic volume}, Galaxies \textbf{2}, 89 (2014).

\bibitem{Hendi2017}
S. H. Hendi, R. B. Mann, S. Panahiyan, and B. Eslam Panah, \textit{Extended phase space thermodynamics and P-V criticality of black holes with a non-linear source}, Phys. Rev. D \textbf{95}, 021501(R) (2017).

\bibitem{Banerjee2012}
R. Banerjee and D. Roychowdhury, \textit{Thermodynamics of phase transition in higher dimensional AdS black holes}, JHEP \textbf{2011}, 4 (2011).

\bibitem{Majhi2017}
B. R. Majhi and S. Samanta, \textit{Thermodynamics and phase transition of a generic higher derivative gravity black hole}, Phys. Lett. B \textbf{773}, 203 (2017).

\bibitem{Li:2016zca}
H.~F.~Li, M.~S.~Ma and Y.~Q.~Ma, \textit{Thermodynamic properties of black holes in de Sitter space,} Mod. Phys. Lett. A \textbf{32} (2016) no.02, 1750017.

\bibitem{Poisson1990}
E. Poisson and W. Israel, \textit{Internal structure of black holes}, Phys. Rev. D \textbf{41}, 1796 (1990).

\bibitem{Bardeen1968}
J. M. Bardeen, \textit{Non-singular general relativistic gravitational collapse}, in Proceedings of the International Conference GR5, Tbilisi, U.S.S.R. (1968).

\bibitem{Hayward2006}
S. A. Hayward, \textit{Formation and evaporation of regular black holes}, Phys. Rev. Lett. \textbf{96}, 031103 (2006).

\bibitem{Ansoldi2008}
S. Ansoldi, \textit{Spherical black holes with regular center}, arXiv:0802.0330 [gr-qc] (2008).

\bibitem{Balart2014}
L. Balart and E. C. Vagenas, \textit{Regular black holes with a nonlinear electrodynamics source}, Phys. Rev. D \textbf{90}, 124045 (2014).

\bibitem{Fan2016}
Z. Y. Fan and X. Wang, \textit{Construction of Regular Black Holes in General Relativity}, Phys. Rev. D \textbf{94}, 124027 (2016).

\bibitem{Lan:2023cvz}
C.~Lan, H.~Yang, Y.~Guo and Y.~G.~Miao, \textit{Regular Black Holes: A Short Topic Review,} Int. J. Theor. Phys. \textbf{62} (2023) no.9, 202.

\bibitem{Singh:2022xgi}
D.~V.~Singh, S.~G.~Ghosh and S.~D.~Maharaj, \textit{Exact nonsingular black holes and thermodynamics,} Nucl. Phys. B \textbf{981} (2022), 115854.

\bibitem{Ma:2014qma}
M.~S.~Ma and R.~Zhao, \textit{Corrected form of the first law of thermodynamics for regular black holes,} Class. Quant. Grav. \textbf{31} (2014), 245014.

\bibitem{Singh:2020xju}
D.~V.~Singh and S.~Siwach, \textit{Thermodynamics and P-v criticality of Bardeen-AdS Black Hole in 4$D$ Einstein-Gauss-Bonnet Gravity,} Phys. Lett. B \textbf{808} (2020), 135658.

\bibitem{Rehan:2024dsg}
M.~Rehan, S.~U.~Islam and S.~G.~Ghosh, \textit{Extended phase space thermodynamics of regular-AdS black hole,} Sci. Rep. \textbf{14} (2024) no.1, 13875.

\bibitem{Singh:2024jgo}
B.~Singh, D.~Veer Singh and B.~Kumar Singh, \textit{Thermodynamics, phase structure and quasinormal modes for AdS Heyward massive black hole,} Phys. Scripta \textbf{99} (2024) no.2, 025305.

\bibitem{Letelier1979}
P. S. Letelier, \textit{Clouds of strings in general relativity}, Phys. Rev. D \textbf{20}, 1294 (1979).

\bibitem{Ghaderi2016}
K. Ghaderi and B. Malakolkalami, \textit{Thermodynamics of Schwarzschild-like black hole with a cloud of strings}, Astrophys. Space Sci. \textbf{361}, 161 (2016).

\bibitem{Toledo2020}
J. M. Toledo and V. B. Bezerra, \textit{Regular black holes with a cloud of strings}, Annals of Physics \textbf{423}, 168349 (2020).

\bibitem{Morais2024}
F. Nascimento, P. H. Morais, J. Toledo, and V. Bezerra, \textit{Thermodynamics and geometry of string cloud spacetimes}, Gen. Relativ. Gravit. \textbf{56}, 86 (2024).

\bibitem{Ma2016}
M.-S. Ma, H.-H. Zhao, L.-C. Cao, and Z.-H. Zheng, \textit{Thermodynamic phase transition of a black hole in a string cloud background}, Int. J. Mod. Phys. A \textbf{31}, 1650120 (2016).

\bibitem{Santos2022}
L. C. N. Santos et al., \textit{Regular black holes in a string cloud background}, Gen. Relativ. Gravit. \textbf{54}, 109 (2022).

\bibitem{Muniz:2025ugk}
C.~R.~Muniz, J.~A.~Rebou{\c{c}}as, L.~T.~de Oliveira, F.~T.~B.~Sampaio and F.~B.~Lustosa, \textit{Regularized black hole solution from a new string cloud source,} Phys. Dark Univ. \textbf{52} (2026), 102272.

\bibitem{Mishra:2026iwq}
V.~K.~Mishra and M.~Pandey, \textit{Thermodynamic Structure of Einstein-Gauss-Bonnet Regular Black Holes Coupled with Cloud of String,} JHAP \textbf{6} (2026) no.2, 87-103.



\bibitem{Singh:2025svv}
D.~V.~Singh, S.~Upadhyay, Y.~Myrzakulov, K.~Myrzakulov, B.~Singh and M.~Kumar, \textit{Thermodynamic behavior and phase transitions of black holes with a cloud of strings and perfect fluid dark matter,} Nucl. Phys. B \textbf{1016} (2025), 116915.
\bibitem{Daassou:2023tmp}
A.~Daassou, R.~Benbrik and H.~Laassiri, \textit{Effect of a cloud of strings and quintessence on a phase transition of charged rotating AdS black holes,} Theor. Math. Phys. \textbf{215} (2023) 893--908.
\bibitem{Okcu2017}
\"{O}. \"{O}kc\"{u} and E. Ayd{\i}ner, \textit{Joule-Thomson expansion of the charged AdS black holes}, Eur. Phys. J. C \textbf{77}, 24 (2017).

\bibitem{Mo2018}
J.-X. Mo, G.-Q. Li, S.-Q. Lan, and X.-B. Xia, \textit{Joule-Thomson expansion of $d$-dimensional charged AdS black holes}, Phys. Rev. D \textbf{98}, 124032 (2018).

\bibitem{Yerra2018}
K. V. Yerra and C. Bhamidipati, \textit{Joule-Thomson expansion of black holes in the extended phase space}, Int. J. Mod. Phys. A \textbf{33}, 1850085 (2018).

\bibitem{Spallucci2013}
E. Spallucci and A. Smailagic, \textit{Maxwell's equal area law for charged Anti-deSitter black holes}, Phys. Lett. B \textbf{723}, 436 (2013).

\bibitem{Lekbich:2023aop}
H.~Lekbich, A.~El Boukili, N.~Mansour and M.~B.~Sedra, \textit{4D AdS Einstein–Gauss–Bonnet black hole endowed with Lorentzian noncommutativity: $P-V$ criticality, Joule–Thomson expansion, and shadow,} Ann. Phys. \textbf{458} (2023) 169451.

\bibitem{Li:2019jcd}
C.~Li, P.~He, P.~Li and J.~B.~Deng, \textit{Joule-Thomson expansion of the Bardeen-AdS black holes,} Gen. Rel. Grav. \textbf{52} (2020) no.5, 50.

\bibitem{Elaima2026}
Y. Elaima, H. Lekbich, A. Daassou, and F. Oubbad, \textit{Rotating Regular Black Hole in a String Cloud Background: Thermodynamics and Shadows}, [Preprint/Accepted in General Relativity and Gravitation] (2026).

\end{thebibliography}
\end{document}